\documentclass[osajnl,twocolumn,showpacs,superscriptaddress,10pt]{revtex4-1} 

\usepackage{amsmath,amssymb,graphicx}
 \usepackage[latin1]{inputenc}
 \usepackage{srcltx}
\usepackage{url}

\begin{document}


\title{Selection of the tagged photons  by off axis heterodyne holography  in Ultrasound-modulated optical
tomography}

\author{  M. Gross}

\address{Laboratoire Charles Coulomb  - UMR 5221 CNRS-UM2
Universit\'e Montpellier
Place Eug\`ene Bataillon
34095 Montpellier, France }


\begin{abstract}
Ultrasound-modulated optical  tomography (UOT) is a technique that
images optical contrast deep inside scattering media. Heterodyne
holography  is a promising tool  able to detect the UOT tagged
photons with high efficiency. In this work, we describe
theoretically the detection of the tagged photon in heterodyne
holography based UOT, show how to filter the untagged photon, and
discuss  the effect of shot noise. The discussion considers also
speckle decorrelation. We show that  optimal  detection sensitivity
can be reached, if the frame exposure time  of the camera used to
perform the holographic detection is of the order of the
decorrelation time.
\end{abstract}


\maketitle
OCIS codes: 170.1650, 170.3660, 290.7050, 090.0090, 170.7050

\section{Introduction}

Light scattering prevents optical imaging deep inside scattering
media.  UOT (ultrasound-modulated optical tomography)
\cite{wang1997ultrasound,elson2011ultrasound} also called
acousto-optic imaging \cite{resink2012state}, has been developed to
overcome this limit by combining ultrasonic defined spatial
resolution and optical contrast (i.e. sensitivity to the bulk
optical properties like absorption).  One of the purpose of the
technique is to use the optical contrast to detect breast tumors
that cannot be seen with ultrasound, because the ultrasound contrast
is too low.
In an UOT experiment,  the light scattered through a diffusing
sample cross an ultrasonic beam, and, due    to the acousto-optic
effect,    undergoes a frequency shift equal to the ultrasonic
frequency \cite{leutz1995ultrasonic,wang2001mechanisms}.
By detecting the  frequency-shifted   photons, called tagged
photons, and by plotting their weight  as a function of the
ultrasonic  beam geometry, 2D (two dimensions) or 3D (three
dimensions) images of the sample can then be obtained with an
ultrasonic spatial resolution.

Various methods have been developed to detect the very low tagged
photon signal out of a large background of untagged photons
\cite{elson2011ultrasound,resink2012state}. First experiments use
single pixel detector   and detection of the tagged photon AC
modulation at the ultrasonic frequency
\cite{wang1995continuous,leutz1995ultrasonic,kempe1997acousto}.
Since each speckle grain oscillates with a different phase, the
single pixel method detects, with a good efficiency, no more than
one speckle grain. This severely limits the detection etendue (the
 etendue is  a property of light in an optical system defined as the product of the detection area and
the acceptance solid angle:  see
\url{https://en.wikipedia.org/wiki/Etendue}).
To increase the detection etendue without reducing the modulation
depth, three types of methods have been developed. The first type
relies on incoherent detection with a narrow spectral filter
($\sim$MHz) that  filter out the untagged light. A large area single
pixel detector can be used. Examples include Fabry-Perot
interferometers
\cite{sakadvzic2004high,kothapalli2008ultrasound,rousseau2009ultrasound}
and spectral hole burning
\cite{li2008detection,li2008pulsed,zhang2012slow} based
methods. These techniques 
require bulky and expensive equipment.
The second and third types of method use    interferences and  are
thus  sensitive to  the  signal phase decorrelation  due to the
living tissue  inner motions, and to the corresponding Doppler
broadening. For breast, this broadening is $1.5$ kHz
\cite{gross2005heterodyne}.
The second method is  based  on a photorefractive  crystal,  which
records the volume hologram of the sample  scattered field. This
hologram can be then used to generate a diffracted  field able to
interfere with  the  scattered field on a large area single-pixel
detector
\cite{murray2004detection,ramaz2004photorefractive,gross2005theoretical,lai2012ultrasound}.
The method has a large optical etendue ($\sim 10^8$ speckle), but is
somewhat sensitive to decorrelation, since the response time of the
crystal is usually much longer than the speckle correlation.
Promising results are expected with Sn2P2S6:Te and Nd:YVO4 crystals,
because of their short response times
\cite{farahi2010photorefractive,jayet2014fast}.

The third type of method  uses a pixel array, i.e., a camera, to
detect the UOT tagged photons
\cite{leveque1999ultrasonic,yao2000frequency,li2002methods,li2002ultrasound}.
The optical etendue ($\sim 10^5$ to $10^6$ speckles) is then related
to the  number of pixels of the camera. The camera method has been
improved by adapting the heterodyne holography technique
\cite{le2000numerical} to the tagged photon detection
\cite{gross2003shot}. By tuning the LO (local oscillator) beam
frequency near the ultrasonic sideband, and by using a properly
adjusted spatial filter,  the tagged photons were  detected
selectively. Moreover, optimal noise detection was obtained,   since
shot noise is the dominant noise in heterodyne holography
\cite{gross2007digital,verpillat2010digital,lesaffre2013noise}.
%

In reference \cite{gross2003shot}, the  UOT tagged photon detection
was  nevertheless performed with a gel phantom sample. The
acquisition time ($\sim 1$ s) was shorter than the gel decorrelation
time ($\sim 10$ s), but much longer than the decorrelation time
\emph{in vivo} ($< 1$ ms). Since \cite{gross2003shot}, it was
generally considered \cite{liu2016lock} that   heterodyne holography
UOT cannot be used with a  sample, whose decorrelation time is
shorter than the time needed to record several camera frames. It
results that heterodyne UOT  has been virtually abandoned.
Very recently,  several groups
\cite{barjean2015fourier,liu2016lock}, have solved this supposed
decorrelation problem by performing heterodyne holography UOT with
a    lock-in camera, which makes  the four phase demodulation very
fast,  within the camera electronics.

\begin{figure}[]
\begin{center}
 \includegraphics[width=6.5cm]{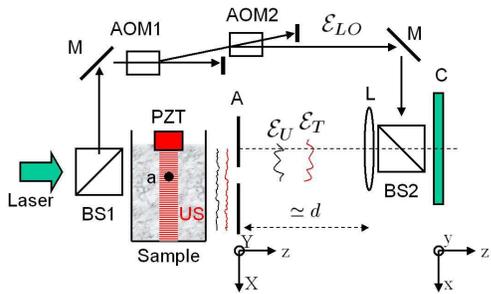}
  \caption{
Typical UOT setup: BS1, BS2: beam splitter;  M: mirror; AOM1,AOM2:
acousto optic modulator; PZT: piezoelectric  transducer that
generates the US (ultrasonic) beam; a: absorber embedded in the
diffusing  sample;  A: rectangular aperture; L: lens;  C: camera;
${\cal E}_{LO}, {\cal E}_{T},{\cal E}_{U}$: LO, tagged and untagged
fields.}\label{Fig_setup}
\end{center}
\end{figure}


In this work,  we show that the decorrelation problem does not
exist. We show that  heterodyne UOT can detect the tagged photon
efficiently  if the decorrelation time is of the order of the
exposure time of a single frame of the camera (which can be order of
magnitudes shorter that the acquisition time of several frames). The
use of a lock-in camera is thus  not essential.

To get this result,  we have developed  a theoretical framework that
describes  the  ref. \cite{gross2003shot}  detection scheme in
detail. We showed  how the untagged photons can be efficiently
filtered  off, and   we calculated how the UOT signal is affected by
untagged photons,  speckle noise,  shot noise,  etendue and
decorrelation ....
By comparing results  obtained with and without decorrelation, we
show that heterodyne holography  remains,   with  decorrelation, an
optimal detection scheme  of the tagged photons. Note that this
point   has been already demonstrated for the detection of the
untagged photon, in experiments done without ultrasound
\cite{atlan2006frequency,atlan2006laser,lesaffre2006effect,atlan2007cortical,atlan2007spatiotemporal,atlan2008high}.

We finally validate our  theoretical analysis by comparing a
theoretical simulation with the experimental results of  reference
\cite{gross2003shot}.
\section{The heterodyne holography UOT setup of ref.\cite{gross2003shot}}

To introduce our theoretical discussion,  let us consider the  heterodyne
holographic UOT  setup of ref. \cite{gross2003shot} (see Fig. \ref{Fig_setup}). 
A laser of frequency $\omega_L$ is split by the  beam splitter BS1
into a signal beam and a local oscillator (LO) beam. The signal beam
travels through the diffusing sample S and is scattered by it. The
sample is explored  by an US (ultrasonic) beam  of  frequency
$\omega_{US}$. The light transmitted by the sample exhibits to
components. The first component  at $\omega_T=\omega_L +
\omega_{US}$ is weak ($\sim 10^{-2}$ to $10^{-4} $ in power), and
corresponds to the tagged  photons  that have interacted with the US
beam.   The second  component  at $\omega_U=\omega_L $ is the main
one ($\simeq 100 \%$ in power). It corresponds to untagged photons
which  have not interacted with  US.
%

A  rectangular aperture A located  off axis  near the sample,
control the size  and location of the zone of the sample where the
tagged and untagged fields ${\cal E}_U$ and ${\cal E}_T$ are
detected.  A lens L of  focal $d$ located at a distance
$|\textrm{AL}|\simeq d$ of A collects the light.
The beam splitter BS2 mixes  ${\cal E}_T$ and ${\cal E}_U$ with the
LO field ${\cal E}_{LO}$ whose frequency $\omega_{LO}$ is controlled
by AOM1 and AOM2 (acousto optic modulator or bragg cell) . To detect
the tagged photons, $\omega_{LO} \simeq \omega_{T} $.

The camera C ($N\times N$ pixels) records a sequence of $M$ frames
$I_m$ (with $m=0...M-1$) corresponding to the   interference
pattern: ${\cal E}_T$ +  ${\cal E}_U$ + ${\cal E}_{LO}$. Frame $I_m$
is recorded at time  $t_m = m \Delta t$, where $\Delta t=2\pi
m/\omega_C$ is the pitch in time, and  $\omega_C$ the camera frame
frequency. The hologram $H_C$ of the aperture A (that is back
illuminated by ${\cal E}_T$ and  ${\cal E}_U$),   in the camera
plane C, is calculated by combining frames $I_m$. The hologram
$H_A$, in the aperture plane A, is then calculated from $H_C$. The
signal of interest (tagged photon) is calculated from $H_A$.

\section{Detection of the tagged photons}

\subsection{Principle of UOT detection}\label{section Principle of detection}
\begin{table}
  \centering
  \begin{tabular}{|l|c|r|}
  \hline
   Heterodyne detection &  Filtering in time:  $\Delta
\varphi \gg 1$\\
   \hline
  $\omega_{LO} \simeq \omega_{T}$  & tagged photon are selected \\
 $\omega_{LO} \ne \omega_{T}$  &  untagged photons are filter off \\
  ${\cal E}_{LO}$ large &  gain: G=$|{\cal E}_T {\cal E}_{LO}|/ |{\cal E}_T|^2 \gg1 $  \\

\hline
   Holographic image of  A  &  Filtering in  space \\
   & $|x_i| > |x_i-x_o|$ and $|x_o| < |w/2|$\\
  \hline
   A off axis &  tagged signal is off-axis\\
    & while untagged signal is on-axis \\
  A narrow &  tagged, untagged  and LO\\
  &  shot noise signals are separated\\
  \hline
\end{tabular}
  \caption{Detection of the tagged photons.}\label{Table_1}
\end{table}

The UOT detection is illustrated by table \ref{Table_1}. The goal is
to measure the energy of the tagged photons $|{\cal E}_T|^2$ which
is very low. To detect selectively the tagged photons, UOT perform a
double filtering (in time and space), with gain.

The time filtering is made by the camera, which records the tagged,
untagged and LO interference pattern. The camera signal is thus
$|{\cal E}_T + {\cal E}_T+ {\cal E}_{LO}|^2$. It results that ${\cal
E}_T$ is detected by heterodyne detection. The tagged photons are
selected (${\cal  E}_T {\cal E}_{LO} $ varies slowly since
$\omega_{LO}\simeq \omega_{T}$), while the untagged photons are
filtered off, (${\cal  E}_U {\cal E}_{LO} $ varies fast since
$\omega_{LO}$ and $\omega_{T}$ are very different). This time
filtering is characterized by the phase drift $\Delta \varphi$ of
the untagged photons during the measurement time. We have: $\Delta
\varphi \simeq  M \omega_{US} \Delta t \gg 1$.

Since the local oscillator field ${\cal E}_{LO} $ is much larger
than the tagged photon field ${\cal E}_T$,  the heterodyne detection
(whose signal corresponds to ${\cal E}_T \cal{E}_{LO} $) is made
with heterodyne gain of $G=|{\cal E}_T {\cal E}_{LO}| / |{\cal
E}_T|^2 \gg 1 $.

The space filtering is made by the aperture A, whose image is
reconstructed by digital holography. Indeed, the camera signal is
also a hologram of the aperture, which is back illuminated by the
tagged field ${\cal E}_T$. By reconstructing the holographic image
of the aperture, one can separate the tagged, tagged LO noise
signals. The tagged photons correspond to the image of the aperture,
which is a bright band located off axis, the untagged photons yield
a parasitic signal located on axis, while the shot noise is a
background which is spread out everywhere : see Fig. \ref{Fig_fig4}
(b) \cite{gross2003shot}. This spatial filtering is characterized by
3 parameters: $x_i$ and $x_o$ and $w$. $x_i$ and $x_o$ are  the
inner and outer coordinates of the aperture edges (with respect to
the detection optical axis), and $w=N \Delta X$ is the width of the
holographic reconstructed image in plane A. Here,  $\Delta X$, given
by Eq.\ref{Eq_Delta_X}, is the size of the reconstructed pixels, and
$N \times N $ is the number of pixels. To perform an efficient
spatial filtering, the aperture A must be within the holographic
reconstructed image: $|x_o| < |w/2|$, and off-axis enough (with
respect to its width): $|x_i| > |x_i-x_o|$.

\subsection{Outline of the UOT simulation}\label{section Principle of the numerical simulation}

Our goal is to analyse  the signals that  are obtained in the
heterodyne UOT experiment of   Fig.~\ref{Fig_setup} in order to
understand how the the tagged photon signal can be detected
selectively.  Our analyse is made  in several steps.

(a)~We assume first that the  tagged and untagged  speckle fields
scattered by the sample in the plane of the aperture A are known.
These fields, which are random, are calculated within  plane A.

(b)~We  calculate the tagged  and untagged  fields in the camera
plane C by field propagation from A to C.

(c)~We assume then that the LO field,  which is flat field, is
known.   We calculate, for each frame $m$, the  intensity $I_m$
corresponding to the sum of the tagged, and untagged and LO fields
on each pixel of the camera, and we convert the optical signal into
photo electrons.

(d) We add to the photo electron signal $I_m$ of each frame and each pixel, a random noise corresponding to shot noise, 
which is the dominant noise in heterodyne holography
\cite{gross2007digital,verpillat2010digital,lesaffre2013noise}. We
get $I'_m$.

We then model the data analysis made in the  UOT experiment of
reference \cite{gross2003shot}.

(e)~We consider all frames $I'_m$ of the sequence, and we calculate
the hologram $H_C$ of the light scattered by the sample.

(f)~We propagate $H_C$ from the camera plane C to the aperture plane
A yielding the reconstructed hologram $H_A$.

(g)~We select within  $H_A$ the tagged photon signal  and to calculate its weight.


\subsection{Notations}

Let us first define the notations for the LO, untagged and  tagged
fields in planes A and C. We have:
\begin{eqnarray}
  {\cal E}_{C,LO}(x,y,t) &=&   E_{LO}~e^{j \omega_{LO} t} + \textrm{c.c. }\\
\nonumber   {\cal E}_{A,U}(X,Y,t) &=&   E_{A,U}(X,Y,t)~e^{j \omega_L t} + \textrm{c.c. }\\
\nonumber   {\cal E}_{A,T}(X,Y,t) &=&   E_{A,T}(X,Y,t)~e^{j (\omega_L+\omega_{US}) t}+ \textrm{c.c. } \\
\nonumber   {\cal E}_{C,U}(x,y,t) &=&   E_{C,U}(x,y,t)~e^{j \omega_L t} + \textrm{c.c. }\\
\nonumber  {\cal E}_{C,T}(x,y,t) &=&   E_{C,T}(x,y,t)~e^{j (\omega_L+\omega_{US}) t} + \textrm{c.c. }
\end{eqnarray}
where $\textrm{c.c.}$ is the complex conjugate. Here, ${\cal
E}_{...}$ are optical fields (which evolve at the optical
frequencies $\omega_L$, $(\omega_L++\omega_{US})$ or $\omega_{LO}$),
while $E_{...}$ are complex amplitudes (which are slow varying with
time). $X,Y$ are the coordinates  in plane A, and $x,y$ the
coordinates in plane C. We have  used different notations for plane
A and C, because the pitches of the calculation grid are different
in plane A and C. To simplify theory, we have considered here that $
E_{LO}$ is flat field, and propagates along the $z$ direction. $
E_{LO}$ do not depend thus on $x,y$ and $t$.

\subsection{The fields in plane A without and with decorrelation}

\begin{figure}
\begin{center}
  \includegraphics[width=3.6cm]{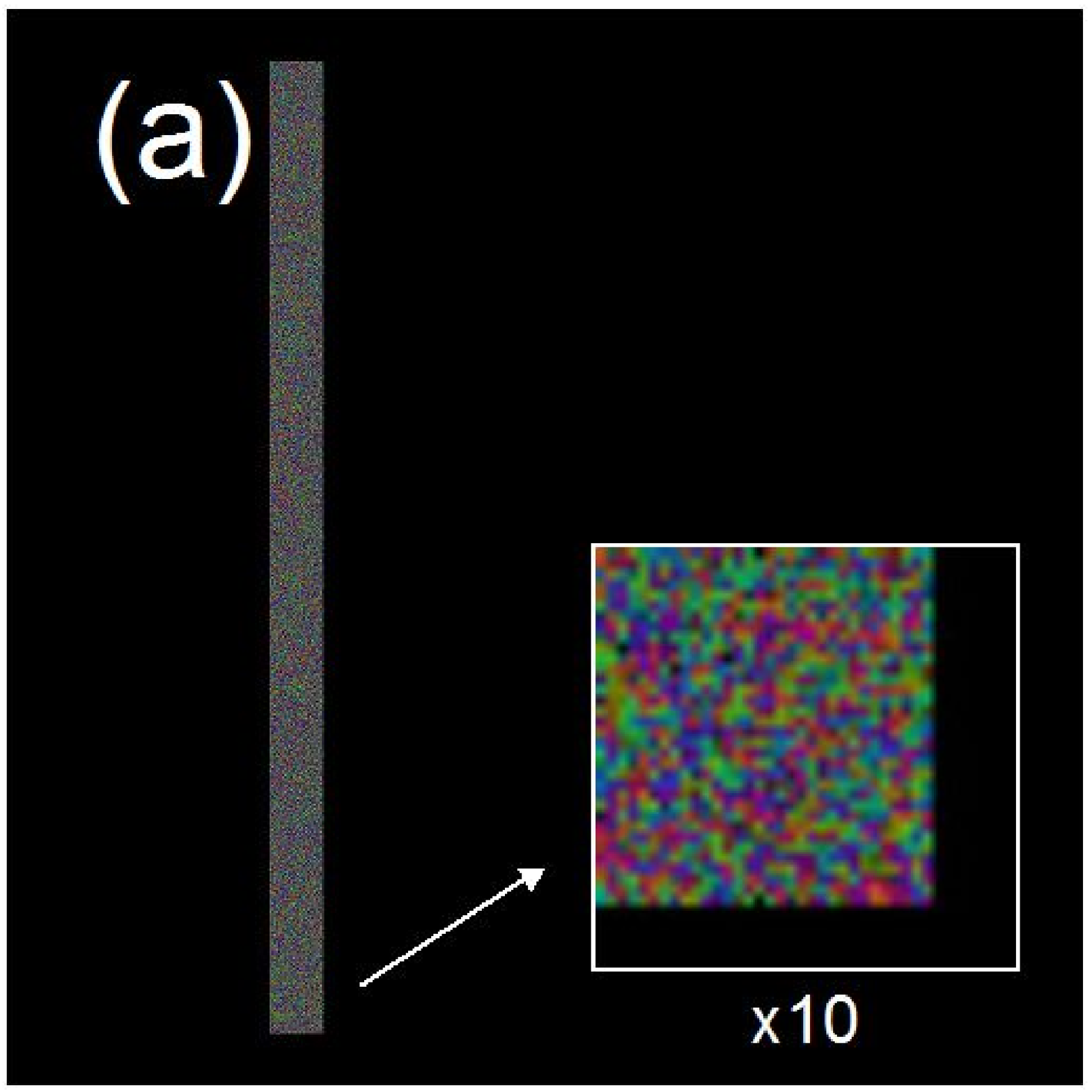}
 \includegraphics[width=3.6cm]{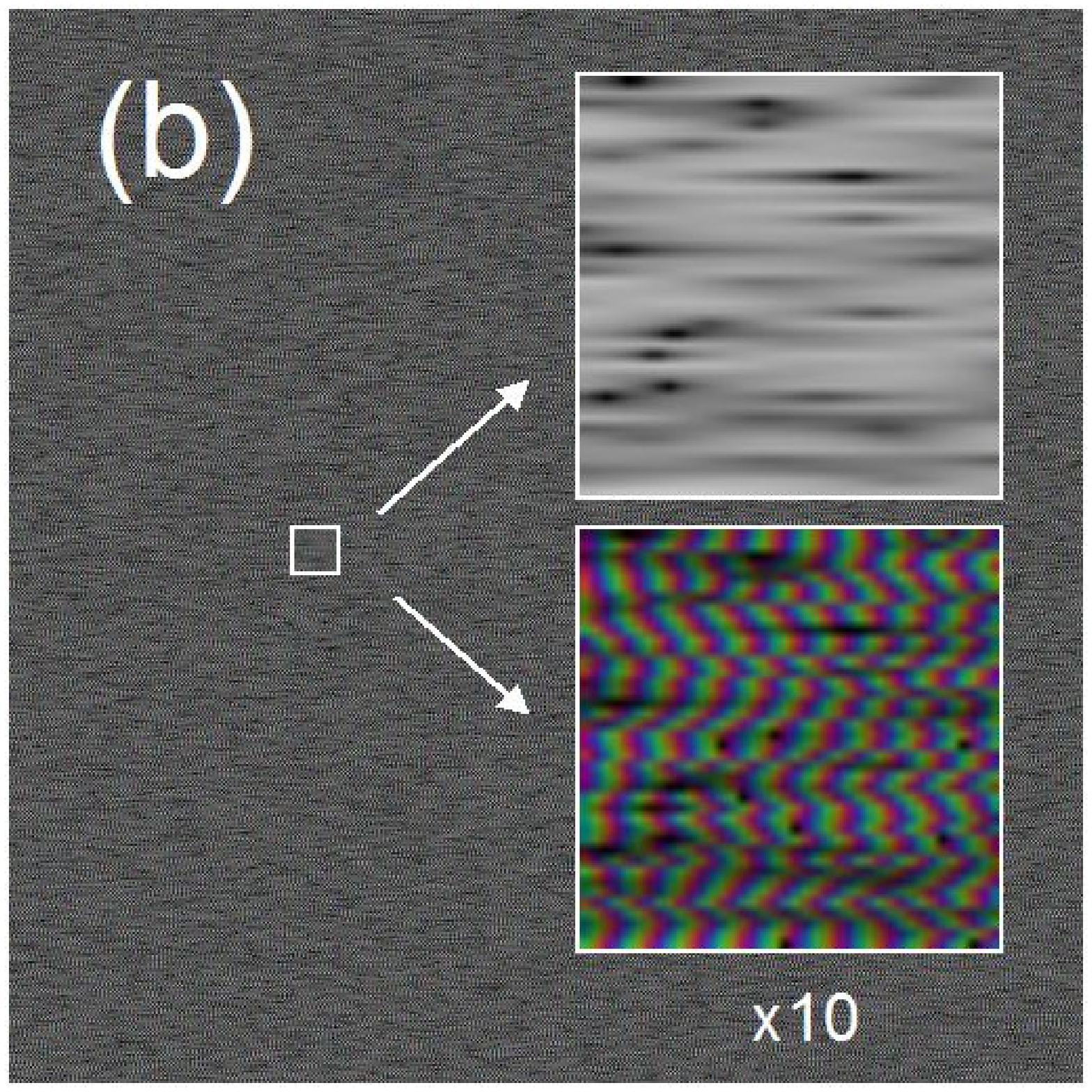}
 \includegraphics[height=3.6cm]{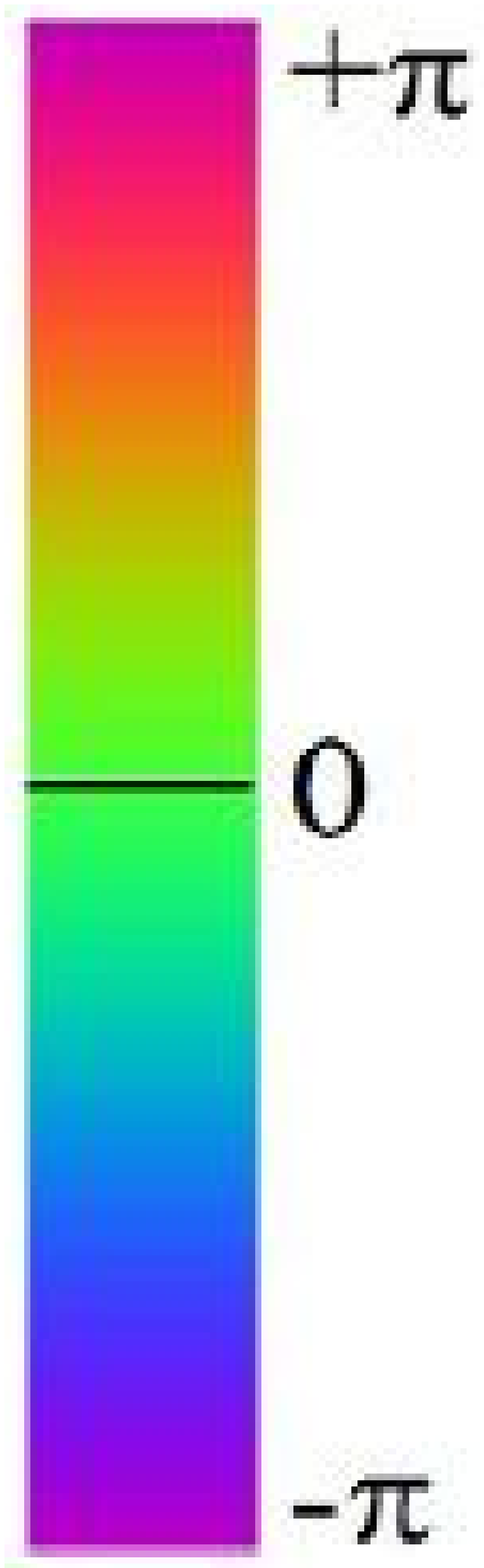}\\
  \caption{
 (a) Image and $\times 10$ zoom (insert) of the
tagged or untagged fields  in the aperture plane plane A i.e.
$E_{A,T}(X,Y,t_m)$ or $E_{A,U}(X,Y,t_m)$.  Brightness is amplitude
in arbitrary logarithmic scale. Color is phase. The coordinates of
the upper left and bottom right corner of the aperture are (250,50)
and (300,974). The calculation grid is  $1024\times 1024$ pixels.
(b) Image and $\times 10$ zoom (insert) of the  tagged or untagged
fields   in the camera plane plane C i.e.  $E_{C,T}(x,y,t_m)$ or
$E_{C,U}(x,y,t_m)$. The upper zoom is amplitude alone in arbitrary
logarithmic scale for intensity.  }\label{Fig_fig2}
\end{center}
\end{figure}

In plane A, the   tagged and untagged photon fields are  fully
developed  speckle. The complex fields are  $ E_{A,T}(X,Y,t_m)$ and
$ E_{A,U}(X,Y,t_m)$ are thus random Gaussian complex quantities that
are uncorrelated from one pixel $(X,Y)$ to any other $(X',Y')\ne
(X,Y)$. These fields are calculated by Monte Carlo on a discrete
calculation grid with  $N\times N$ pixels of pitch $\Delta X$.
Because of the aperture A, the fields $ E_{A,T}(X,Y,t_m)$ and $
E_{A,U}(X,Y,t_{m})$ are zero out of the aperture.
\begin{enumerate}
  \item
If the experiment is made with a gel sample,  whose speckle remains
correlated with time, the random amplitudes $ E_{A,T}(X,Y,t_m)$ and
$ E_{A,U}(X,Y,t_m)$ do not depend on the recording time $t_m$ of
frame $m$.
  \item
On the other hand, if the experiment is  made with a living sample,
whose scatterers move, we will consider that the speckles are
totally decorrelated from one frame to the next.  The random  fields
$ E_{A,T}(X,Y,t_m)$ and $ E_{A,U}(X,Y,t_m')$ are then totally
uncorrelated if $t_m \ne t_{m'}$.
\end{enumerate}

Figure \ref{Fig_fig2} (a) shows an  example of tagged or untagged
fields   $E_{A,T}(X,Y,t_m)$ or $E_{A,U}(X,Y,t_m)$ calculated by
Monte Carlo. As seen in the $\times 10$ zoom, the phase is random
from one pixel to the next.

\subsection{The tagged and untagged fields in plane C}

In the UOT setup of Fig.\ref{Fig_setup}, the lens L, which is
located near the camera, and whose focal plane is close to plane A,
collects the tagged and untagged  fields. Because of lens L, the
tagged and untagged fields in planes C and A are related by a
Fourier transform
\begin{eqnarray}\label{Eq_propag_A2C}
  E_{A,T}(X,Y) &=& {\tilde E_{C,T}}(k_x,k_y)=  \textrm{FFT} (E_{C,T}(x,y))\\
 \nonumber  E_{A,U}(X,Y) &=& {\tilde E_{C,U}}(k_x,k_y)= \textrm{FFT} (E_{C,U}(x,y))
\end{eqnarray}
where FFT is the discrete Fourier transform operator. The
coordinates $(X,Y)$ in plane A are related to the Fourier space
coordinates $(k_x,k_y)$ by:
\begin{eqnarray}
 (X,Y)= (k_x,k_y) \times |\textrm{CA}|/k
\end{eqnarray}
where $|\textrm{CA}|$ is the camera to aperture distance and  $k=2
\pi/ \lambda$.

Figure \ref{Fig_fig2} (b)  shows the  tagged or untagged fields
$|E_{C,T}(x,y,t_m)|^2$  or  $|E_{C,U}(x,y,t_{m})|^2$ in the camera
plane. The image exhibits a speckle that is uniformly distributed on
the calculation grid that corresponds to the camera detector area.
To better visualize this  speckle $\times 10$ zooms  are displayed
in the square inserts. Since the aperture is vertical, the speckles
are elongated in the  horizontal direction $x$ (see upper insert),
but because of the off axis location of the aperture, the phase
varies very fast, and  increases   along the $x$ direction, within
each speckle. Indeed,  in the lower insert the colors are displayed
in green, red, blue order from left to right.

\subsection{Pixel, etendue, modes and camera low pass filtering }

In  equation \ref{Eq_propag_A2C}, the FFT is calculated    within a
calculation grid that fits with the camera pixels. The pitch $\Delta
x$ of the discrete coordinates $x,y$ is thus equal to the size of
the pixel of the camera. Because of the FFT, the   pitch $\Delta X$
in plane A is
\begin{eqnarray}\label{Eq_Delta_X}
    \Delta X&=& 2 \pi |\textrm{CA}| ~  /( N k \Delta x  )
\end{eqnarray}
where $N \times N$ is the number of pixels of the camera ($N=1024$ typically). The detection etendue $G$ is thus
\begin{eqnarray}\label{Eq_eq5}
G &=& S_A  S_D/ |\textrm{CA}|^2\\
\nonumber &=&N^2\lambda^2
\end{eqnarray}
where  $S_A=N^2|\Delta X|^2$ and $S_C=N^2|\Delta y|^2$  are the
areas of the calculation grid in plane A and C.

The number of modes or  speckle grains that can be detected is equal
to  the number of pixels of the camera:  $N^2$, i.e.  to the number
of pixels of the calculation grid in plane A and C. Equation
\ref{Eq_eq5} means that the etendue of each pixel of planes A, or C
is equal to $\lambda^2$, i.e. to the etendue of one mode. We must
nevertheless notice that the fields in plane A ($E_{A,T}$ and
$E_{A,U}$) vary very fast in the $X$ and $Y$ directions. Indeed,
because these fields results from the scattering by a think
diffusing sample, their correlation length is about $\lambda$. The
spatial variations of these fields are thus considerably faster than
the pitch $\Delta X$ of the measurement grid.

To solve this paradox, we must remark that  the  camera plays  the
role of a low pass filter that selects the slow varying  components
(in space) of the fields in plane A. Indeed, the fields $E_{A,T}$
and $E_{A,U}$  are scattered in all directions, and most of the
scattered  photons never reach the camera. The photons, which do not
reach the camera correspond to the fast varying components of
$E_{A,T}$ and $E_{A,U}$. The camera area is a low pass filter that
selects the low spatial frequency components of the fields $E_{A,T}$
and $E_{A,U}$, which are sampled  by pixels of size $\Delta X \gg
\lambda$. The fields $E_{A,T}$ and $E_{A,U}$ that are considered
correspond to these slow components.

This analysis is confirmed by  the energy conservation in planes A
and C, since we have, because of the FFT Percival relation:
\begin{eqnarray}
  \sum_{X,Y} |E_{A,U \textrm{or} T}(X,Y)|^2 &=&  \sum_{x,y} |E_{C,U \textrm{or} T}(x,y)|^2
\end{eqnarray}

\subsection{The camera frame signals $I_m$}

The frame signal $I_m$ corresponds to the  sum of the tagged,
untagged and  LO fields. To detect the tagged photons, $\omega_{LO}$
is made close to the tagged photon frequency $\omega_L+\omega_{US}$.
The  tagged photons field $E_T$  thus interfere with $E_{LO}$, and
$E_T$ and $E_{LO}$ must be summoned in field. On the other hand, the
untagged photon field $E_U$ does not interfere with $E_{LO}$ and
$E_T$. $E_U$ can be summoned in intensity. We have thus:
  \begin{eqnarray}\label{Eq_Im}
        I_{m}(x,y) &=&  \left|E_{C,T}(x,y,t_m ) + c^m  E_{LO} \right|^2  \\
        \nonumber &&   ~~~~~+ \left|E_{C,U}(x,y,t_m )\right|^2\\
    \nonumber     &=&  \left(c^mE_{LO}^* E_{C,T}(x,y,t_m) + ~\textrm{c.c.} \right)\\
       \nonumber &&   ~~~~~ +  |E_{LO}|^2+  \left|E_{C,T}(x,y,t_m )\right|^2 \\
          \nonumber &&   ~~~~~ +  \left|E_{C,U}(x,y,t_m )\right|^2
  \end{eqnarray}
where   $c$ is the LO versus tagged field phase shift that
correspond to one time step $\Delta t$:
 \begin{eqnarray}
c=e^{j(\omega_{LO}-\omega_{US}-\omega_{L})\Delta t}
\end{eqnarray}
%
%
In equation \ref{Eq_Im}, the useful terms  that  are  $ c^m E_{LO}^*
E_{C,T}$  and $ c^{-m} E_{LO} E^*_{C,T}$. These  terms   are
enhanced because of the high power of the local oscillator $E_{LO}$
(i.e. the  holographic gain  $|E_{LO}^* E_{C,T}|/ |E_{C,T}|^2$ is
much larger than one).

The term $ c^m E_{LO}^* E_{C,T}$  is the $+1$ grating order term,
which proportional to $E_{C,T}$. This term  is  displayed on
Fig.\ref{Fig_fig2}(b). Similarly, the term $ c^{-m} E_{LO}
E^*_{C,T}$ is the $-1$ grating order term. The terms  $|E_{LO}|^2$
and $\left|E_{C,T}\right|^2$ are zero grating order terms.
$|E_{LO}|^2$ is  large, but flat field, while
$\left|E_{C,T}\right|^2$ is small, and can be neglected. Finally,
the term $\left|E_{C,U}\right|^2$ is an untagged photon spurious,
which is most often much larger than the useful term $ c^m E_{LO}^*
E_{C,T}$, because $\left|E_{C,U}\right|^2 \gg
\left|E_{C,T}\right|^2$. This spurious must be filtered off.

\subsection{The shot noise}

Because of the random nature of light emission  and camera  photo
conversion, the frame optical signal $I_m$ is affected by  shot
noise yielding the frame detected signal $I'_m$:
\begin{eqnarray}\label{Eq_I'_m}
 I'_m(x,y)&=&  I_{m}(x,y) + s(x,y,m)\sqrt{I_{m}(x,y)} ~~)
\end{eqnarray}
where the term $s\sqrt{I_{m} }$ accounts  for shot noise. Here,
$I_m$ must be expressed   in photo electron Units per pixel and per
frame, while $s$ is a real Gaussian random variable of variance $
\langle s^2\rangle=1$ uncorrelated with pixels (i.e. $X,Y$)  and
with frames (i.e. with $m$).
%

\subsection{The holograms in plane A and C and the selection of the tagged photons
}

The tagged photon signal  can be extracted  from the measured data
by calculating  the hologram $H_C$  in the camera plane C, and by
propagating $H_C$ from the camera plane C to the aperture plane A
yielding $H_A$. The way this procedure is done depends on the
decorrelation of the speckle.

We will  show first how $H_C$ and $H_A$ are  calculated without and
with decorrelation (see section  \ref{section_calc_without_decorr}
and section \ref{section_calc_with_decorr}). To illustrate how the
tagged photons are selected without and with decorrelation, we will
consider examples, in which the shot noise is neglected (see section
\ref{section_example_without_decorr} and section
\ref{section_example_with_decorr}).

\subsubsection{Calculation of the holograms  without
decorrelation}\label{section_calc_without_decorr}

Let us first consider a sample whose speckle  remains correlated
with time. The holograms in plane C and A are thus
$H_{C,\textrm{corr}}$ and  $H_{A,\textrm{corr}}$. The recorded data
are analyzed by four phase detection at the tagged photon frequency.
We have
\begin{eqnarray}\label{Eq_LO_corr}
 \omega_{LO}= \omega_{LO,\textrm{corr}} = \omega_L+ \omega_{US}+ \omega_C/4
\end{eqnarray}
 yielding $c=-j$ in Eq.~\ref{Eq_Im}.
The hologram $H_{C,\textrm{corr}}$ is calculated from a sequence of $M$ frames, where $M$ is a multiple of four.
We have: 
\begin{eqnarray}\label{Eq_corr}
    H_{C,\textrm{corr}}(x,y)&=&\sum_{m=0}^{M} j^m I'_m(x,y)
\end{eqnarray}
The   hologram  $H_{A, \textrm{corr}}$  is calculated from $H_{C,\textrm{corr}}$ by Fourier transform:
\begin{eqnarray}\label{Eq_H_A}
H_{A,\textrm{corr} }(X,Y) &=& {\tilde H_{C,\textrm{corr}}}(k_x,k_y)\\
\nonumber &=&   \textrm{FFT}\left(H_{C,\textrm{corr}}(x,y)\right)
\end{eqnarray}

\subsubsection{Example of hologram calculated without decorrelation and without shot noise}\label{section_example_without_decorr}

To illustrate how the tagged photons can be selected,   let us
consider that the shot noise can be neglected. Let us assume $I'_m =
I_m$.

With this hypothesis,  the terms  $|E_{LO}|^2$,
$\left|E_{C,U}\right|^2$ and $\left|E_{C,T}\right|^2$ of
Eq.\ref{Eq_Im} do not contribute to $H_{C,\textrm{corr}}$, because
they do not vary with  time $t_m$. Moreover, the order $-1$ term  $
c^{-m} E_{LO} E^*_{C,T}$ yields zero in the summation of
Eq.\ref{Eq_corr}, because $j^m c^{-m}=-1^m$. Finally, the $+1$ term
contributes alone. We have:
\begin{eqnarray}\label{Eq_corr_bis}
   H_{C,\textrm{corr}}(x,y)& \simeq &M E^*_{LO} E_{C,T}(x,y)\\
   \nonumber &\propto& E_{C,T}(x,y)
\end{eqnarray}
The hologram  $H_{C,\textrm{corr}}$ is thus proportional to the
field  $E_{C,T}$,  which is displayed on Fig.\ref{Fig_fig2}.
From  Eq.\ref{Eq_propag_A2C},  Eq.\ref{Eq_corr_bis} and  Eq.\ref{Eq_H_A}, we get:
\begin{eqnarray}
H_{A, \textrm{corr}}(X,Y)&\simeq& M E^*_{LO} E_{A,T}(X,Y)\\
\nonumber &\propto&  E_{A,T}(X,Y)
\end{eqnarray}
The reconstructed   hologram $H_{A, \textrm{corr}}$ is thus
proportional to the tagged field $E_{A,T}$. The image displayed in
in Fig. \ref{Fig_fig2}(a) corresponds thus both to $E_{A,T}$ and to
$H_{A, \textrm{corr}}$.

\subsubsection{Calculation of the holograms  without
decorrelation}\label{section_calc_with_decorr}

Let us now consider a living sample whose speckle is fully
decorrelated from one frame to the next. The holograms, in planes C
and A, are thus  $H_{C,\textrm{decorr}}$ and
$H_{A,\textrm{decorr}}$. The recorded data are  analyzed by two
frames  detection at the tagged photon frequency, without phase
shift. Indeed, the  phase shift is irrelevant, because the phase is
lost, from one frame to the next. We have:
\begin{eqnarray}\label{Eq_LO_decorr}
 \omega_{LO}= \omega_{LO,\textrm{decorr}} = \omega_L+ \omega_{US}
\end{eqnarray}
yielding  $c=1$ in Eq.~\ref{Eq_Im}.
The hologram $H_{C,\textrm{decorr}}$ is calculated by:
\begin{eqnarray}\label{Eq_decorr}
   H_{C,\textrm{decorr}}(x,y)&=&I_0(x,y)-I_1(x,y)
\end{eqnarray}
The reconstructed hologram   $H_{A, \textrm{decorr}}$  is calculated
from $H_{C,\textrm{decorr} }$ by Fourier transform:
\begin{eqnarray}\label{Eq_H_A_decorr}
 H_{A,\textrm{decorr} }(X,Y) &=& {\tilde H_{C,\textrm{decorr}}}(k_x,k_y)\\
\nonumber &=&   \textrm{FFT}(~H_{C,\textrm{decorr}}(x,y))
\end{eqnarray}
Because of Eq.\ref{Eq_decorr},  the hologram
$H_{C,\textrm{decorr}}$  is real (phase is zero or $\pi$). The
reconstructed hologram $H_{A,\textrm{decorr}}(X,Y)$, obtained by
Eq.\ref{Eq_H_A_decorr}, is thus symmetric with respect to
$(X,Y)=(0,0)$.

\subsubsection{Example of hologram calculated without decorrelation
and without shot noise}\label{section_example_with_decorr}

Here again, let us let us neglect the shot noise and assume  $I'_m =
I_m$.

With this hypothesis,  the terms $|E_{LO}|^2$  and
$\left|E_{C,T}\right|^2$ of Eq.\ref{Eq_Im}  do not   contribute to
$H_{C,\textrm{decorr}}$, because   $|E_{LO}|^2$ do not vary with
time, and because $\left|E_{C,T}\right|^2$  can be neglected. We
have:
\begin{eqnarray}\label{Eq_eq18}
 ~\nonumber H_{C,\textrm{decorr} }(x,y)  &\simeq& \sum_{m=0}^1 (-1)^m  \left(E_{LO}^* E_{C,T}(x,y,t_m) + ~\textrm{c.c.} \right)\\
  &&~~~~+  \sum_{m=0}^1 (-1)^m \left|E_{C,U}(x,y,t_m)\right|^2
\end{eqnarray}
Here,    $\left(c^mE_{LO}^* E_{C,T}(x,y,t_m) + ~\textrm{c.c.}
\right)$ is a tagged photon term, which  can be used to extract the
tagged photon signal. On the other hand,    $\left|E_{C,U}\right|^2$
is  a zero order terms, related to the untagged photons. This term
must be filtered off.

\begin{figure}
\begin{center}
  \includegraphics[width=3.6cm]{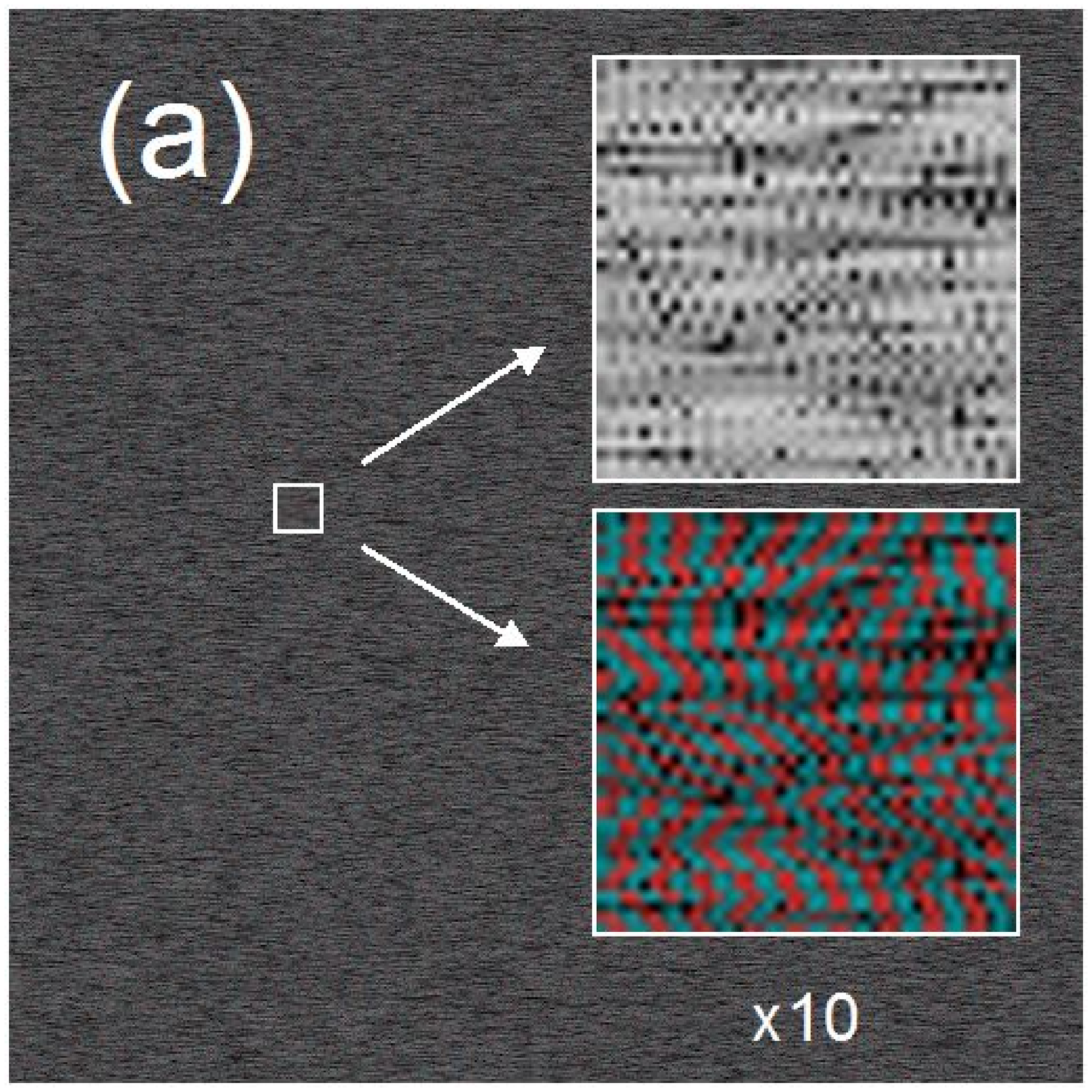}
 \includegraphics[width=3.6cm]{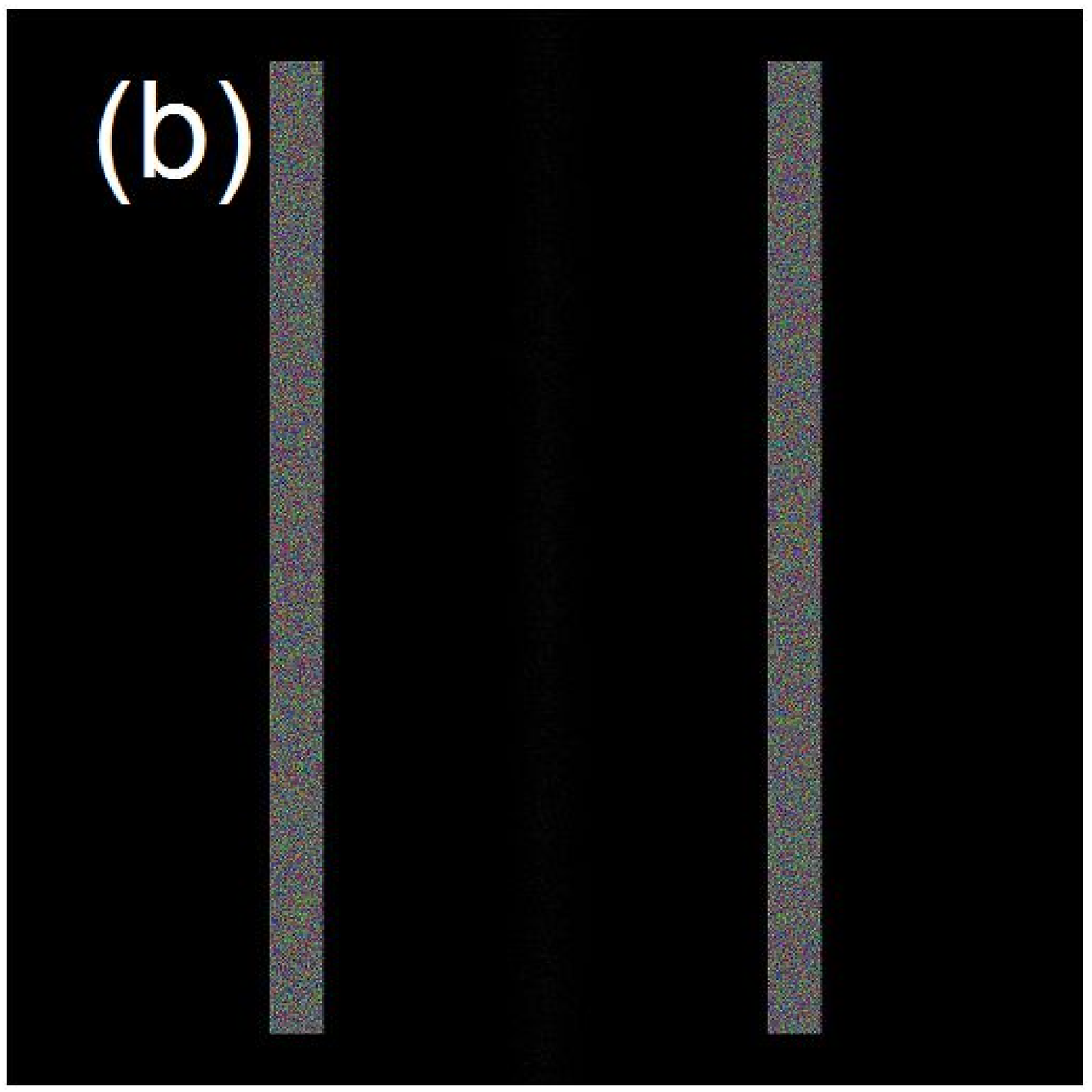}
 \includegraphics[width=3.6cm]{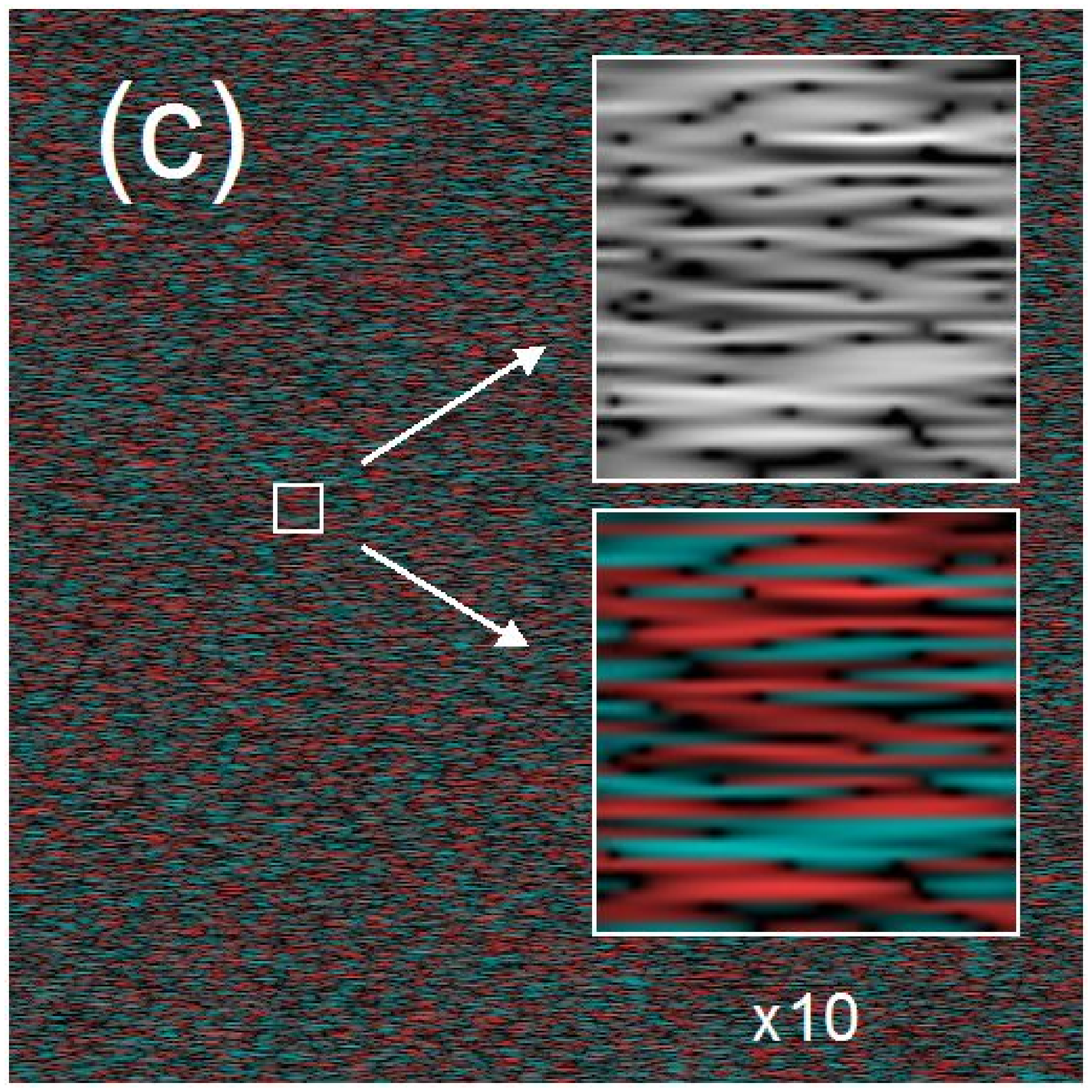}
  \includegraphics[width=3.6cm]{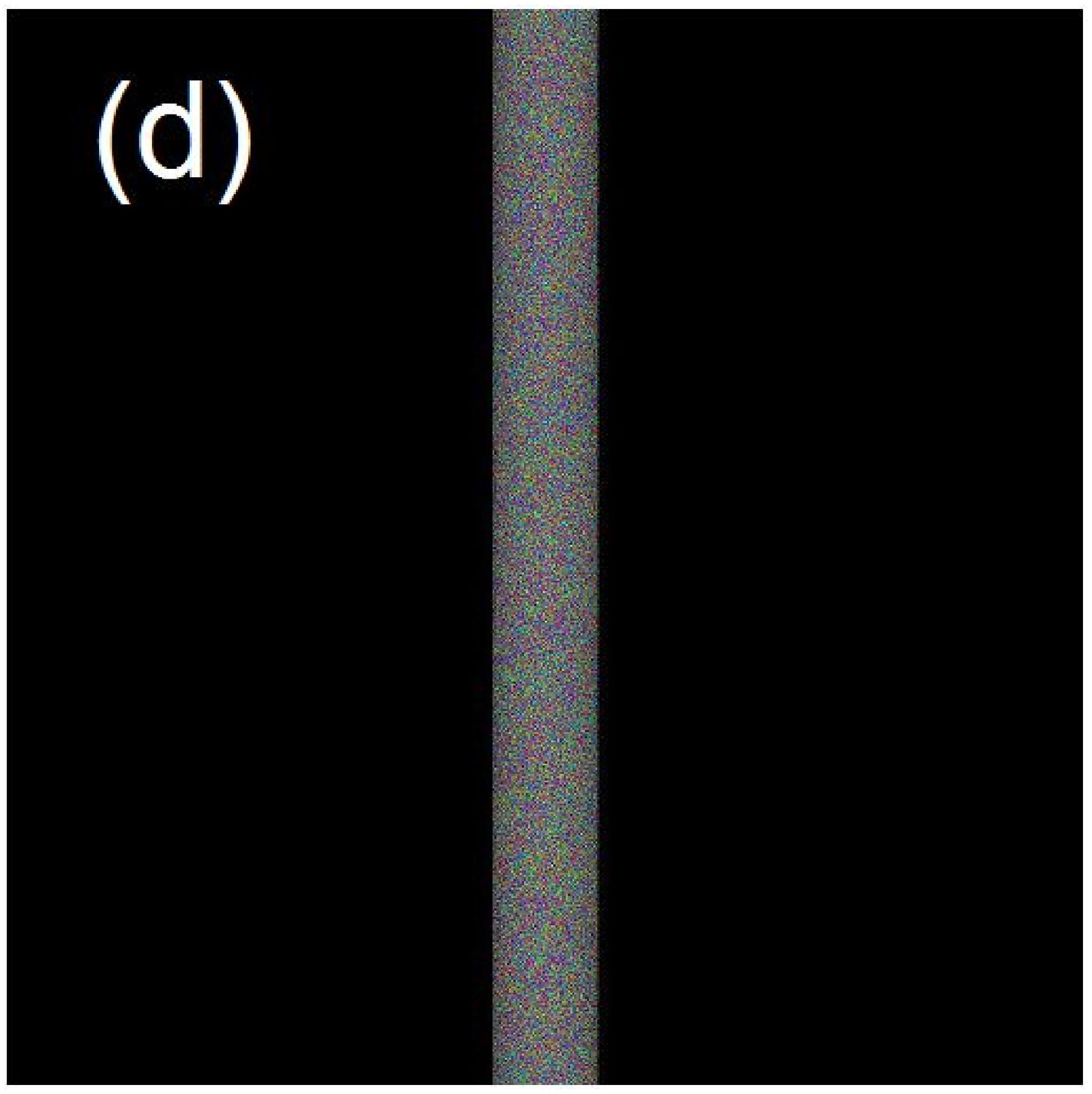}
 \caption{
(a,c) Images and $\times 10$  zoom (inserts) of the holograms
$H_{C,\textrm{decorr}}(x,y)$   calculated with decorrelation for the
tagged (a) or untagged (c) photons alone. (b,d) Images of the
holograms   $H_{A,\textrm{decorr}}(X,Y)$ reconstructed with tagged
(b) and untagged (d) photons alone.   Brightness is the amplitude in
arbitrary logarithmic scale. Color is the phase.  The upper zooms in
(a) and (c) are displayed with amplitude alone.
   }\label{fig_decorr_holo}
\end{center}
\end{figure}

To illustrat how the tagged and untagged photons  can be separated,
we have calculated  the holograms $H_{C,\textrm{decorr}}$ and
$H_{A,\textrm{decorr}}$ with tagged or  untagged photons alone.
\begin{enumerate}
  \item
With tagged photons alone,  $ (c^mE_{LO}^* E_{C,T}(x,y,t_m) +
~\textrm{c.c.} )$ contributes alone in Eq.~\ref{Eq_eq18}. Because
the aperture A is off axis,  $I_0$, $I_1$ and thus
$H_{C,\textrm{decorr}}$, which is real, varies very  fast with $x$:
see Fig. \ref{fig_decorr_holo}~(a). It results that the
reconstructed hologram  $H_{A,\textrm{decorr}}$ calculated by FFT
(see Eq.~\ref{Eq_H_A_decorr}) is zero in the center of the
calculation grid, and exhibits both the $+ 1$ (left bright
rectangle)  and $-1$ (right rectangle) grating order images of the
aperture: see Fig.\ref{fig_decorr_holo}~(b).
\item
With untagged  photons alone,  the term $ \left|E_{C,U}\right|^2 $
contributes alone.   Because the aperture A is thin,   $I_0$, $I_1$
(and thus $H_{C,\textrm{decorr}}$) vary slowly with $x$: see Fig.
\ref{fig_decorr_holo} (c). It results that the  reconstructed
hologram $H_{A,\textrm{decorr}}$  is zero except  in the center of
the calculation grid, where it  exhibits a bright band of signal:
see Fig. \ref{fig_decorr_holo} (d). The width of this band is  twice
the width of the rectangles of Fig.\ref{fig_decorr_holo}~(b).
\end{enumerate}
The tagged and untagged  contribution to the reconstructed hologram
$H_{A,\textrm{decorr}}$ are thus located in different regions of the
calculation, and can be thus easily separated.

\section{ The detection sensitivity   without an  with decorrelation}

%
\subsection{The Units for the  energy of the fields }

To simplify the analysis, the  tagged and untagged photon energies
in plane A are quantified in Units of photo electron per  pixel, and
per $T_C$, where $T_C$ is defined by the following way.
\begin{enumerate}
  \item
Without decorrelation,  $T_C$ is the recording time of the sequence
of $M$ frames.  $T_C$  is shorter than the decorrelation time.
  \item
With decorrelation, $T_C$ is the exposure time of one frame.  $T_C$
is of the order of the decorrelation time, but much shorter that the
camera time pitch $\Delta t$.  The fields $E_{A,T}$ and $E_{A,U}$
remain thus correlated during the exposure time $T_C$, but are fully
uncorrelated from one frame to the next.
\end{enumerate}

In experiments, the  local oscillator field $E_{LO}$ can be freely
adjusted. The best results are obtained by adjusting  $E_{LO}$ to be
as large as possible without saturating the camera
\cite{gross2007digital,verpillat2010digital,lesaffre2013noise}. We
have considered a camera, whose saturation level is about $2\times
10^4$ photo electrons, and a local oscillator power that corresponds
to half saturation, i.e. $|E_{LO}|=10^4$  photo electrons per pixel
of plane C, and per frame.

In typical heterodyne  UOT experiments, the tagged photon field
$E_{A,T}$ is very low. It corresponds to about one photon electron
per pixel  or less. On the other hand, the untagged photon field
$E_{A,U}$ is much larger than $E_{A,T}$, with typically
$|E_{A,U}|^2/|E_{A,T}|^2 \sim 10^3$. Since the tagged photon signal
is very low, it is essential to account for shot noise.
%

\subsection{ The reconstructed  hologram $H_A$}

To simulate the UOT experiment,  we first   calculated the tagged
and untagged speckle fields  $E_{A,T}(X,Y,t_m)$ and
$E_{A,U}(X,Y,t_m)$ in plane A by  Monte Carlo. We considered here a
$1024\times 1024$ calculation grid and an aperture, whose upper left
and bottom right corner coordinates are $(X,Y)=(125, 0)$ and
$(300,1023)$. These parameters correspond to $x_i=-212$, $x_o=-387$,
 and $w=1024$ in $\Delta X$ Units. We made the
following assumption  for the  statistical averages
$\langle|E_{A,U}|^2 \rangle$, and $\langle|E_{A,T}|^2\rangle$ of the
field energies   $|E_{A,U}(X,Y,t_m)|^2$ and  $|E_{A,T}(X,Y,t_m)|^2$
in plane A.
\begin{enumerate}
  \item
Without decorrelation,  $M=12$ ($T_C$ is the recording time of 12
frames), $ \langle|E_{A,U}|^2 \rangle= 10^4$ and $\langle|E_{A,T}|^2
\rangle=1$ photo electrons per pixel and per $T_C$.
  \item
With decorrelation,  $ \langle|E_{A,U}|^2 \rangle= 250$ and
$\langle|E_{A,T}|^2 \rangle= 1$  photo electrons per pixel and per
$T_C$.
\end{enumerate}


\begin{figure}
\begin{center}
  \includegraphics[width=4.2cm]{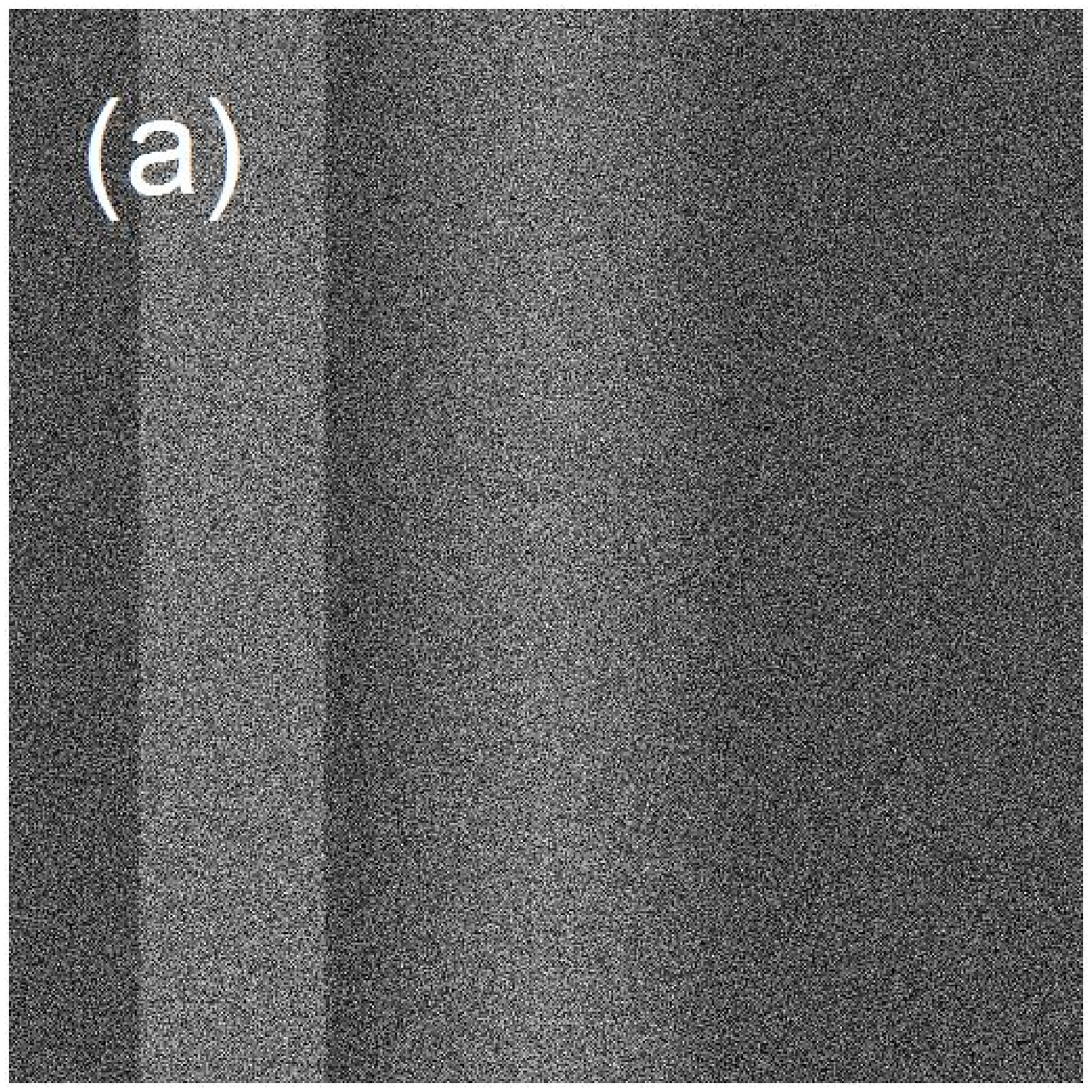}
 \includegraphics[width=4.2cm]{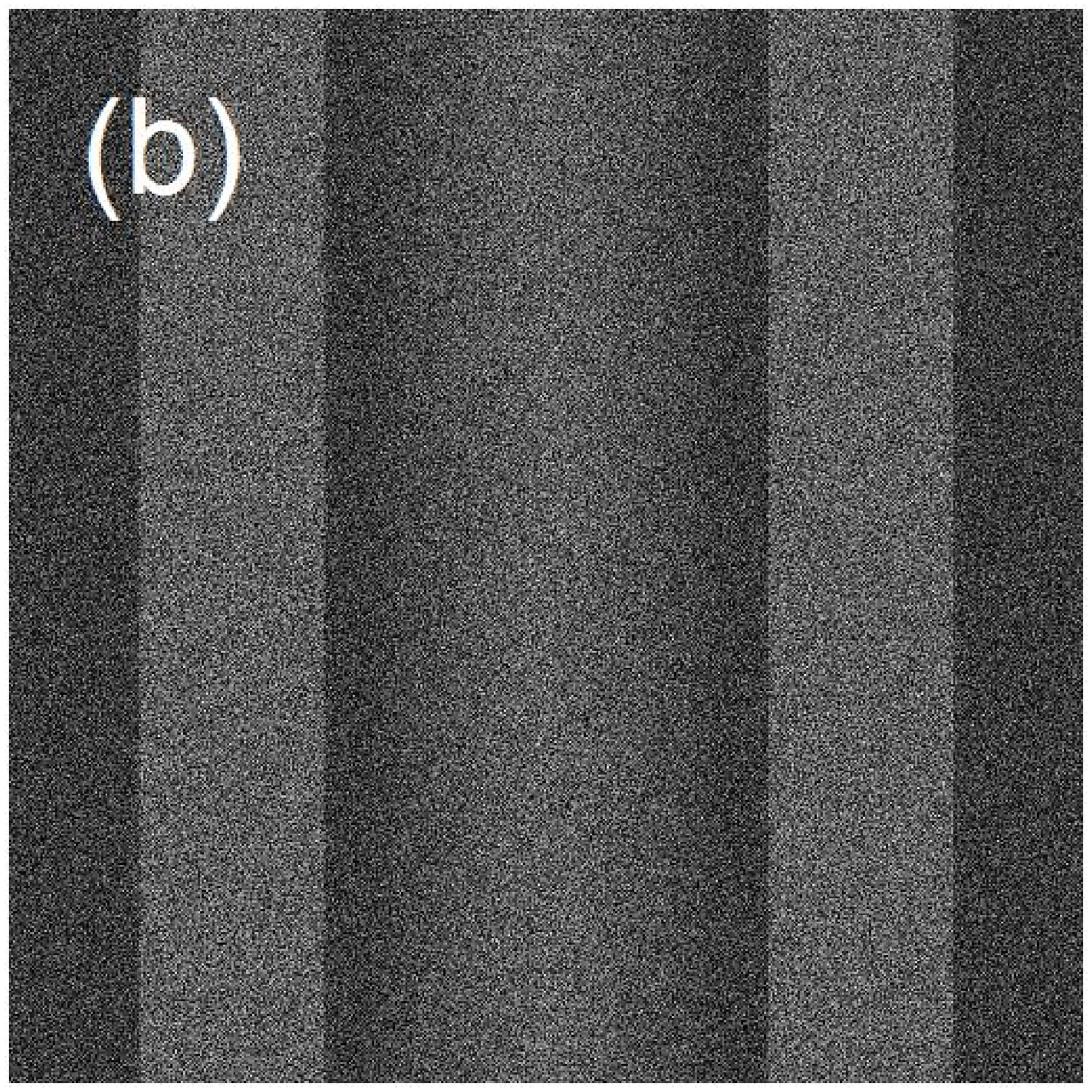}\\
   \includegraphics[width=4.2cm]{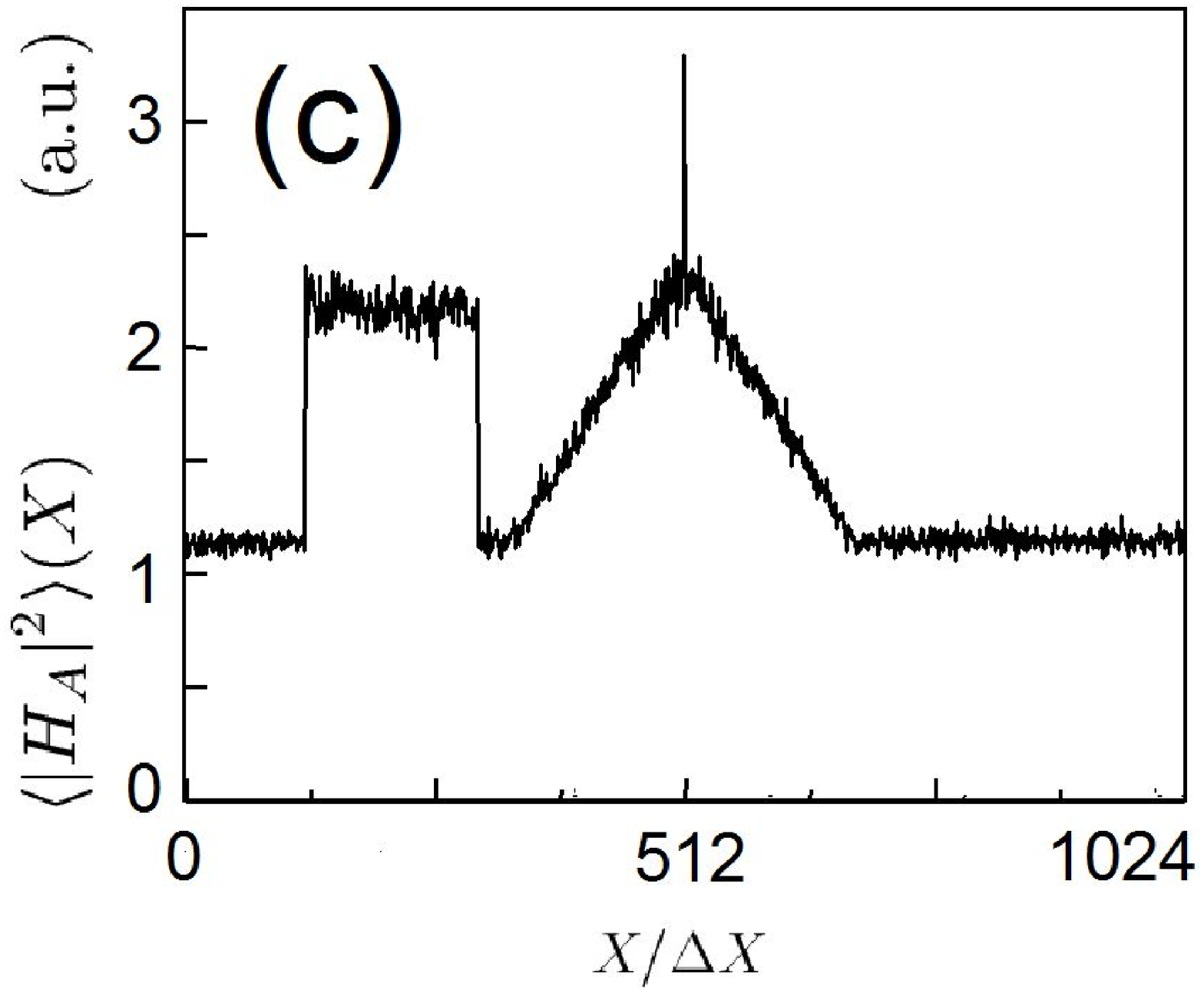}
  \includegraphics[width=4.2cm]{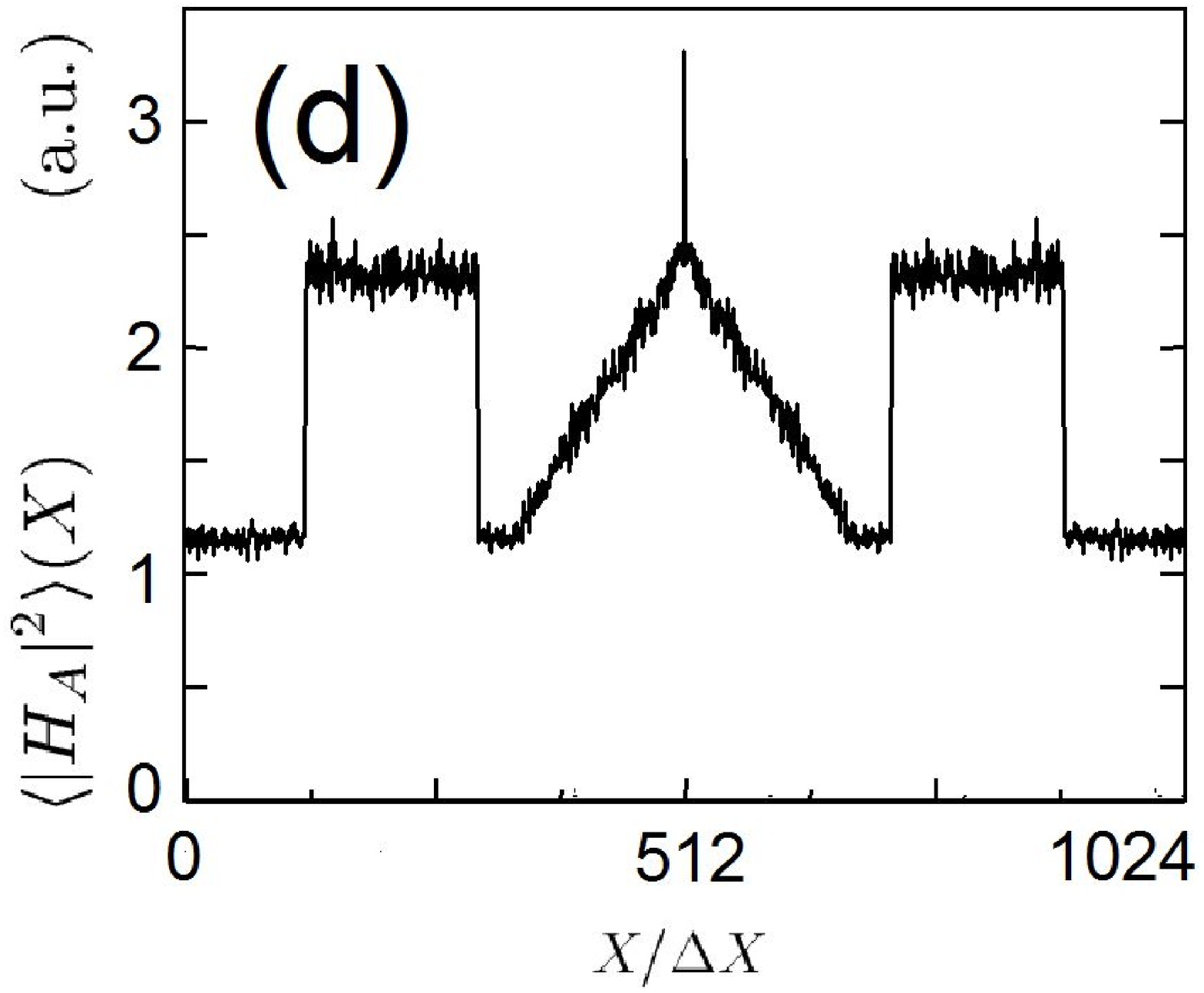}
  \caption{
Holograms  $H_{A}(X,Y)$ (a,b) and curves $\langle  |H_{A}|^2
(X)\rangle$  (c,d)  obtained by calculation with $x_i=-212$,
$x_o=-387$ and $w=1024$ in $\Delta X$ Units. The images (a,b) are
displayed in an arbitrary logarithmic scale for  $|H_{A}(X,Y)|^2$ ,
the curves (c,d) in arbitrary linear scale. }\label{Fig_fig3}
\end{center}
\end{figure}

With these initial conditions, we have calculated  the reconstructed
hologram i.e. $H_{A}=H_{A, \textrm{corr}}$ or $H_{A,
\textrm{decorr}}$, and displayed them in   Fig.
\ref{Fig_fig3}~(a,b).
\begin{enumerate}
  \item
Without decorrelation (Fig. \ref{Fig_fig3}~(a)~), the tagged photon
signal corresponds to the $+1$ image of the aperture, i.e. to the
left hand side bright rectangular zone.
The blurred bright zone, in the center of  Fig. \ref{Fig_fig3} (a),
corresponds to a parasitic detection of the untagged photon signal,
which does not cancel completely here because of  shot noise.
Finally, the shot noise yields a flat noise background in all points
of the images.

  \item
With decorrelation (Fig.~\ref{Fig_fig3}~(b)), the tagged photon
signal corresponds to the $\pm 1$ images of the aperture, i.e. to
the bright rectangular zones located in the left and right hand side
of Fig.~\ref{Fig_fig3}~(b).
The blurred bright zone, in the center of  Fig. \ref{Fig_fig3} (b)
corresponds  to the untagged photon signal, which  is much larger
with decorrelation than without. In order to get the roughly the
same brightness for the untagged photon signal, the calculation has
been done  with a much lower untagged energy with decorrelation
($250$ photo electrons) than without ($10^4$). Here again, the shot
noise yields a flat noise background.
\end{enumerate}

\subsection{ The profile of $H_A$}

To analyse  more quantitatively the   reconstructed hologram
$H_{A}$, we have averaged $|H_{A}(X,Y)|^2 $ over $Y$ to get
$\langle |H_{A}|^2 \rangle(X)$ \cite{gross2003shot,ruan2013pulsed}:
\begin{eqnarray}\label{Eq_profile_HA}
  \langle |H_{A}|^2 \rangle (X) &=& \frac{1}{N}  \sum_Y |H_{A}(X,Y)|^2
 \end{eqnarray}
where $H_{A}=H_{A, \textrm{corr}}$ without decorrelation and $H_{A}=H_{A, \textrm{decorr}}$ with.

Figure \ref{Fig_fig3} (c,d) shows  the curves  $\langle  |H_A(X)|^2
\rangle$   obtained without (c)   and with (d) decorrelation. The
rectangular walls  located on the left hand side of
Fig.~\ref{Fig_fig3} (c) and in the left and right side of
Fig.~\ref{Fig_fig3} (d) correspond to the tagged photon.
On the other hand, the  triangular  walls located on the center of
Fig.~\ref{Fig_fig3} (c) and (d) correspond to the untagged photons.
The width of the triangular walls is twice the width of the
rectangular walls $|x_i-x_o|$, which is equal to the width of the
aperture in $\Delta X$ Units. Out of the rectangular and triangular
walls, the curves exhibit a flat background noise floor that
corresponds to shot noise.

\begin{figure}
\begin{center}
  \includegraphics[width=4.2cm]{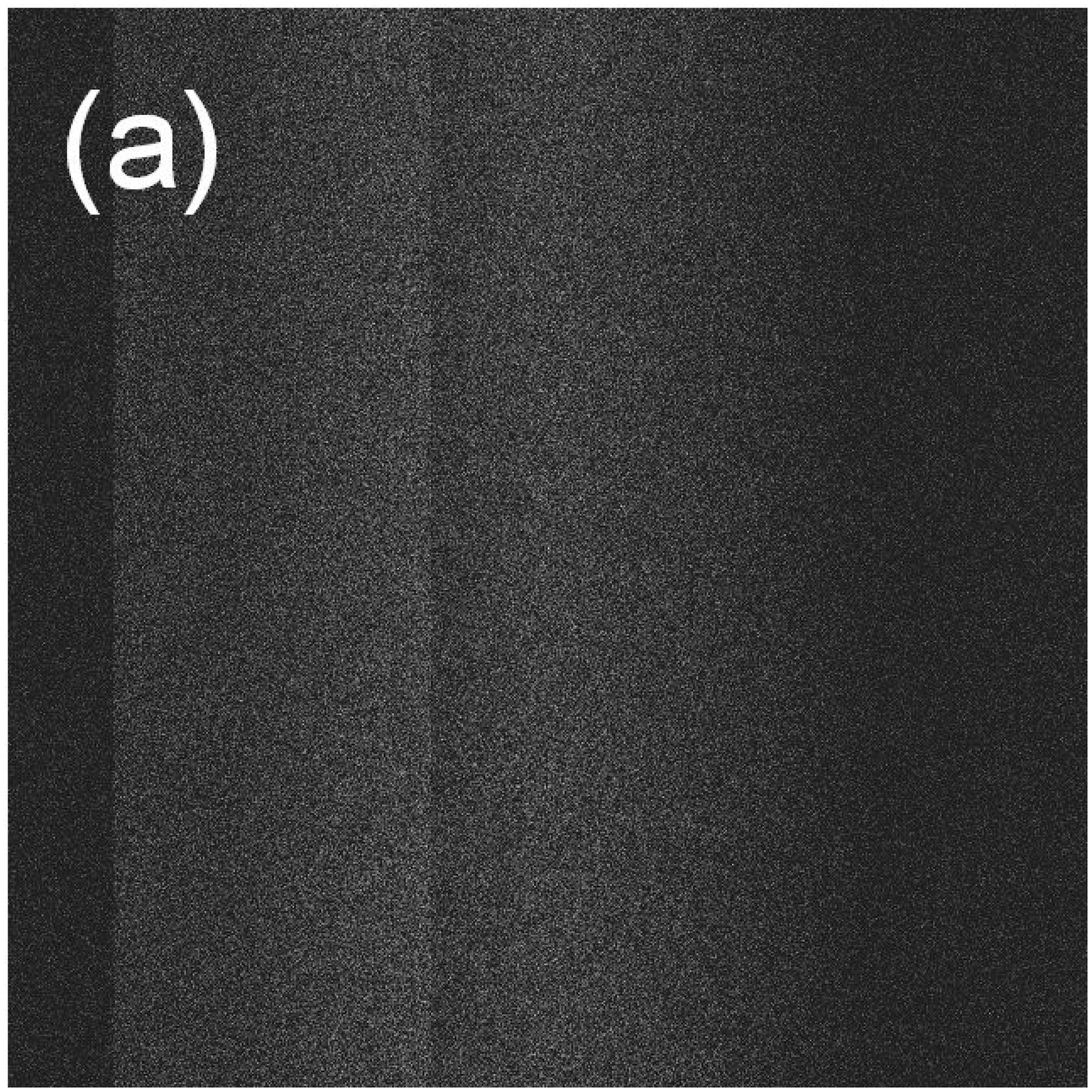}
 \includegraphics[width=4.2cm]{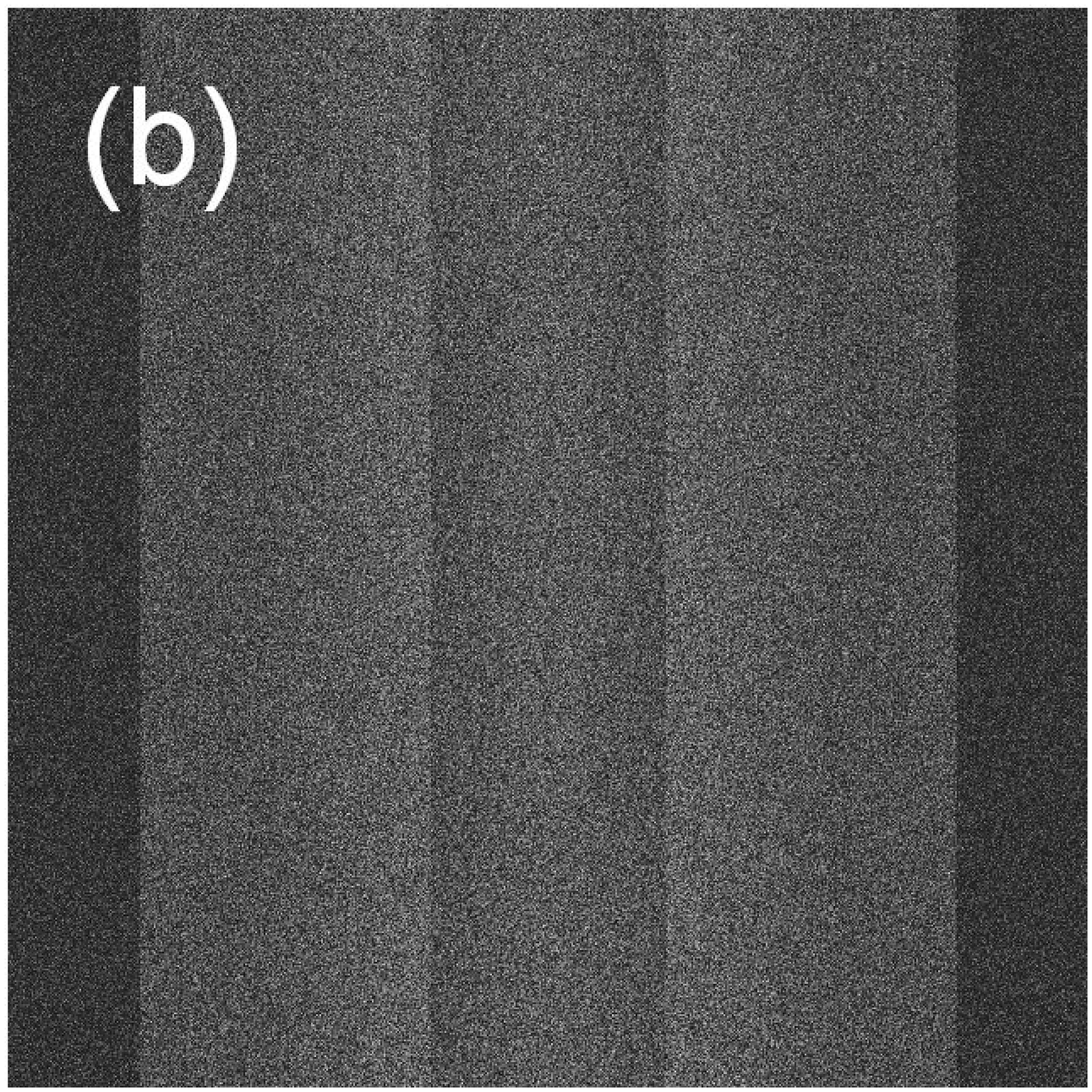}\\
 \includegraphics[width=4.2cm]{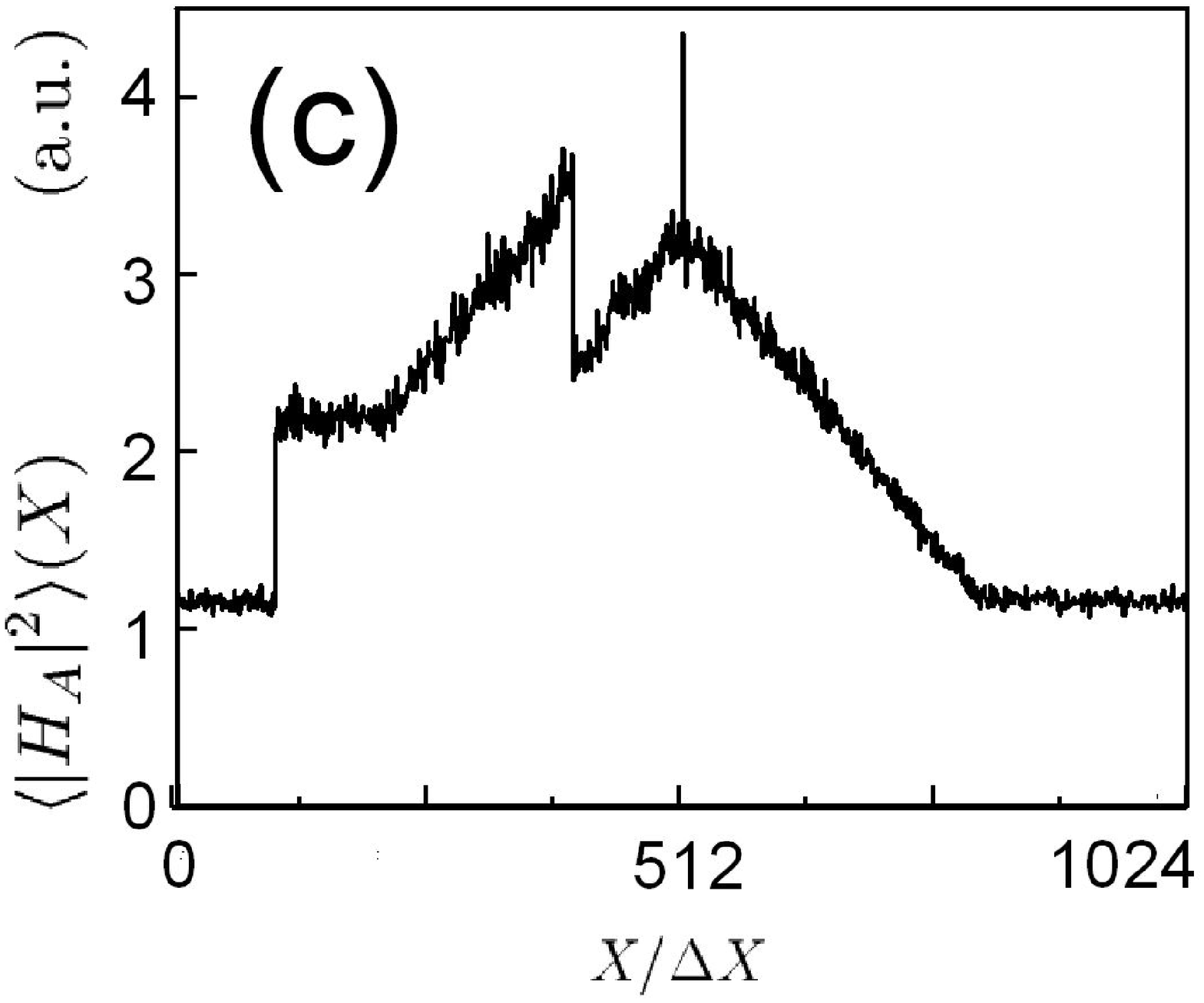}
 \includegraphics[width=4.2cm]{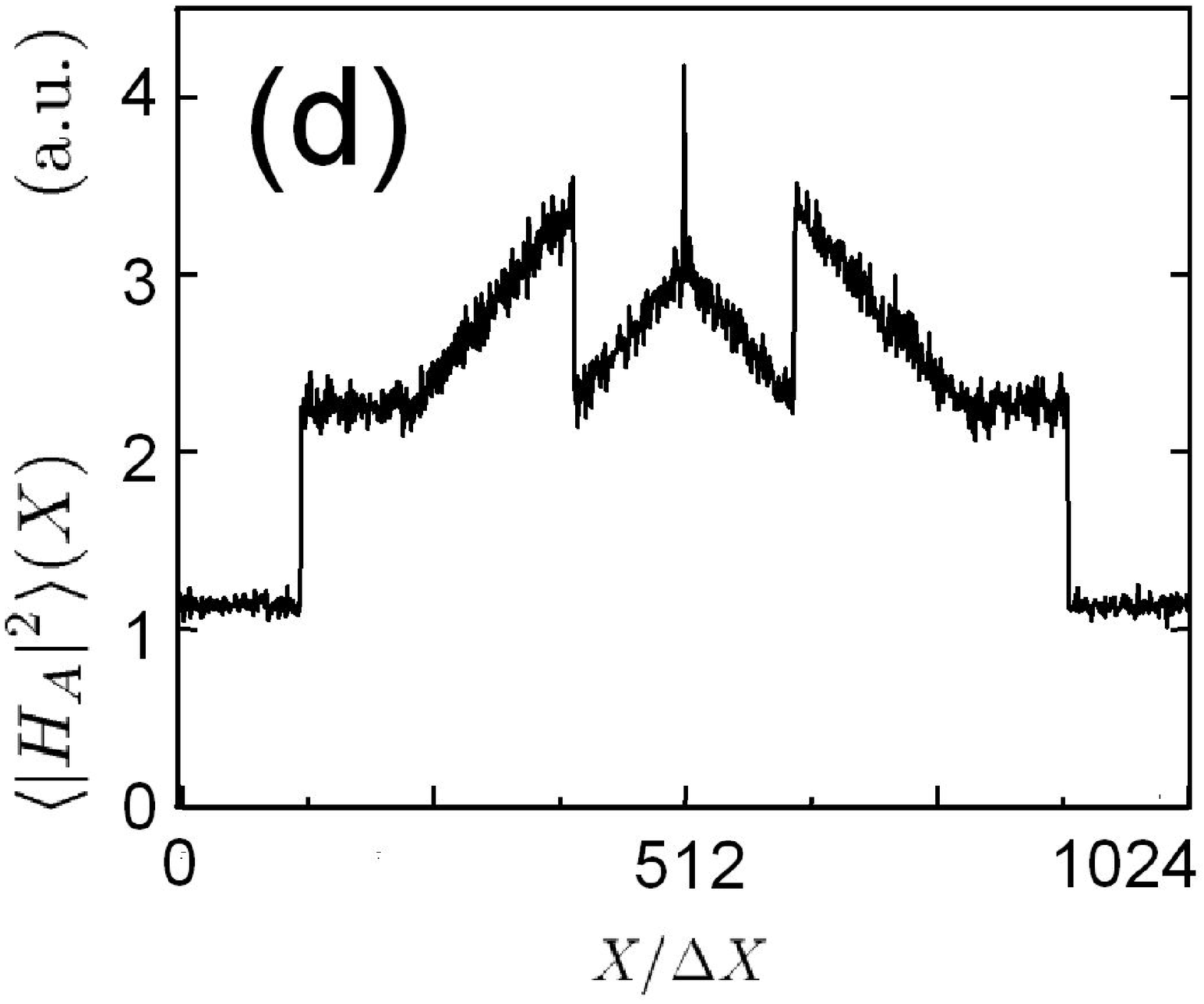}
  \caption{
Holograms  $H_{A}(X,Y)$ (a,b) and curves $\langle  |H_{A}|^2
(X)\rangle$  (c,d)  obtained by calculation with $x_i=-112$,
$x_o=-387$ and $w=1024$ in $\Delta X$ Units. The images (a,b) are
displayed in an arbitrary logarithmic scale for  $|H_{A}(X,Y)|^2$ ,
the curves (c,d) in arbitrary linear scale. }\label{Fig_fig4bis}
\end{center}
\end{figure}

Here, by a proper choice of the aperture width and aperture off axis
position ($x_i=-212$, $x_o=-387$ and $w=1024$ ), that obedience to
the conditions $|x_i|> |x_i-x_o|$ and $|x_o| < w/2$, the rectangular
and triangular walls are well separated, making possible to filter
off the unwanted untagged photons. Here the condition $|x_i|>
|x_i-x_o|$ is needed to separate the tagged and  untagged photon
signals. If this condition is not fulfilled, the two signals are
mixed together as shown by Fig.\ref{Fig_fig4bis}, which is obtained
in the same condition that Fig.\ref{Fig_fig3} but with $x_i=-112$,
$x_o=-387$ and $w=1024$, i.e. with $|x_i|< |x_i-x_o|$. On the other
hand, the condition $|x_o| < w/2$ is needed to image the aperture
properly i.e. without holographic reconstruction aliases.

By obeying to conditions $|x_i|> |x_i-x_o|$ and $|x_o| < w/2$, is
also possible to select zones (like pixels 0 to 124 and 898 to 1023)
without signal, where the measured signal corresponds to shot noise.
Since the height of the tagged photon rectangular walls is equal to
the noise  floor,  the shot noise   equivalent signal is equal to
one  tagged  photo electron per pixel (i.e. per etendue $\lambda^2$)
and per $T_C$, where $T_C$ is either the recording time of the
sequence (without decorrelation), or the exposure time of one frame
(with decorrelation).

%
%

\subsection{The detection sensitivity limits}

To evaluate the tagged photon detection sensitivity limits, we have
calculated, without  and  with  decorrelation,  the $\langle
|H_A|^2\rangle (X)$ curves  by  varying  the tagged   photon energy.
\begin{enumerate}
  \item
Without decorrelation, the calculation is made with $M=12$ frames,
$|E_{LO}|^2= 10^4$,     $ \langle|E_{A,U}|^2 \rangle= 10^4$, and
$\langle|E_{A,T}|^2 \rangle= \alpha$ with $\alpha =1$, 0.5, 0.25 and
0.125  for curves 1 to 4 (in  photo electron per pixel of plane A,
et per $T_C$ Units).
  \item
With decorrelation, the calculation is made $|E_{LO}|^2= 10^4$,  $
\langle|E_{A,U}|^2 \rangle= 250$ and   $\langle|E_{A,T}|^2 \rangle=
\alpha$.
\end{enumerate}

\begin{figure}
  \begin{center}
%
  \includegraphics[width=4.2cm]{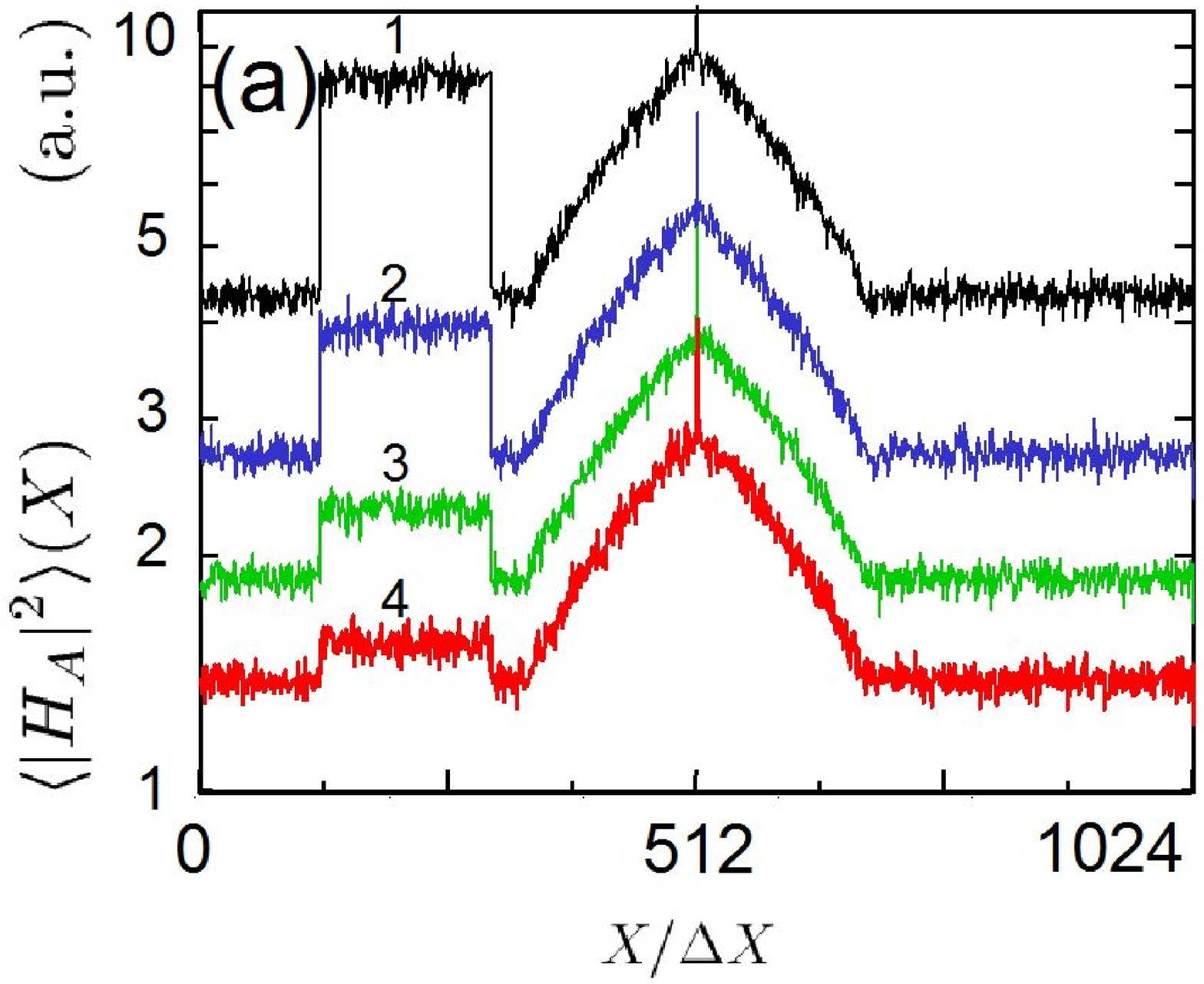}
 \includegraphics[width=4.2cm]{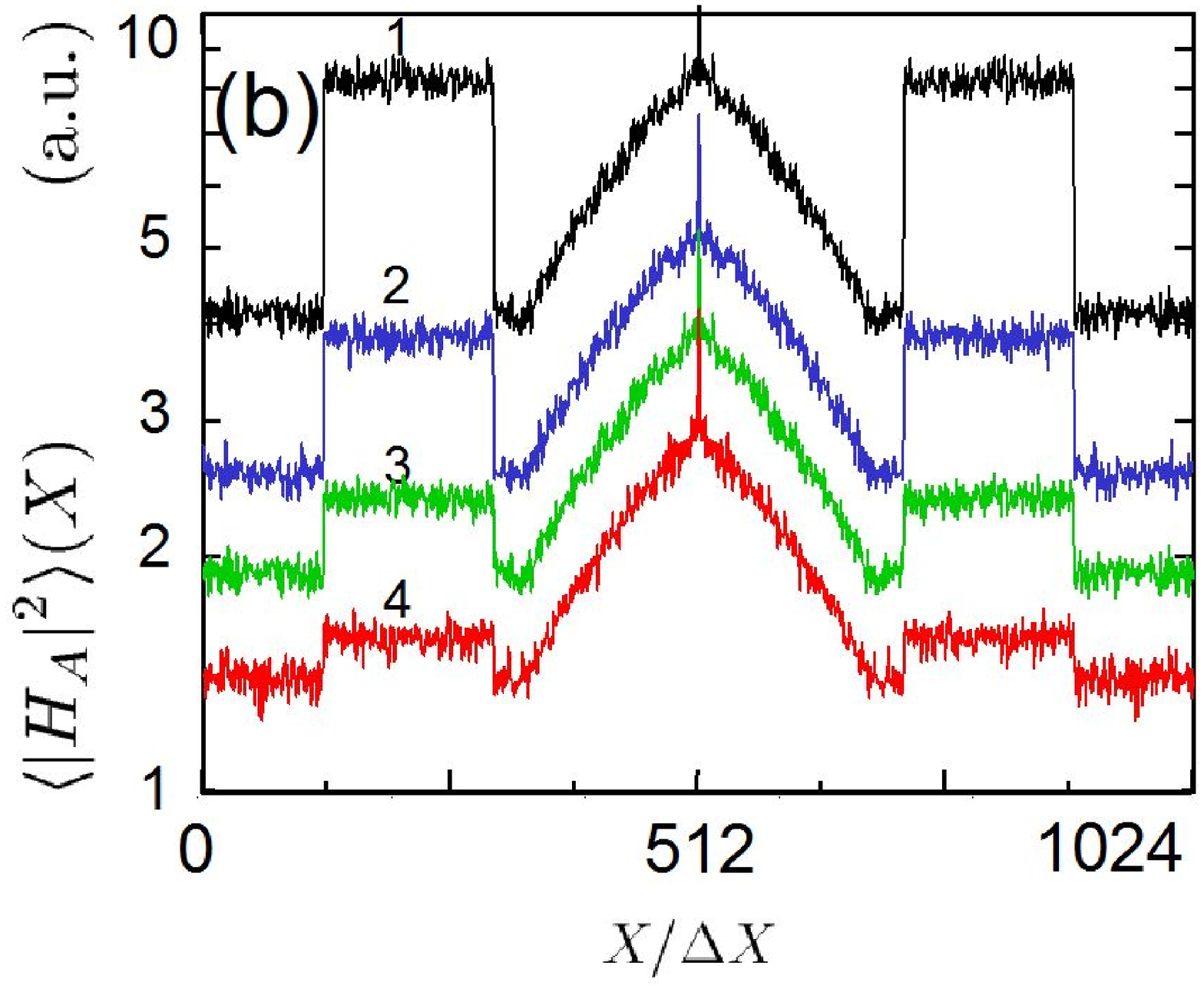}
    \caption{
Curves $\langle  |H_A|^2 \rangle(X)$ obtained without (a) and with
decorrelation (b).  Calculation is made with $\alpha = 1$ (1), 0.5
(2), 0.25 (3) and 0.125 (4), where $\alpha$  is the number of tagged
photons  per pixel and per $T_C$. Plots are   made in arbitrary
logarithmic scale.  }\label{Fig_Fig5}
 \end{center}
\end{figure}

%
%

The curves  $\langle |H_A|^2\rangle (X)$ are drawn in
Fig.\ref{Fig_Fig5}.  To better visualize them, the curves were
plotted  in log scale, and  the curves  were arbitrarily  shifted up
or down to better separate them from each other. Figure
\ref{Fig_Fig5} shows that  we get roughly  the same sensitivity for
the detection of the tagged photon with and without decorrelation.

By averaging over  the about $10^5$ pixels of the rectangular
aperture, the sensitivity limit is improved down to about $\alpha
\sim 1/\sqrt{10^5} \sim 1/300$ photo electron per pixel and per
$T_C$. This result agrees with what observed experimentally for the
detection of the untagged photons \cite{gross2005heterodyne}.

\section{Validation of the theory with experiment \cite{gross2003shot} }

\begin{figure}
\begin{center}
  \includegraphics[width=4.2cm]{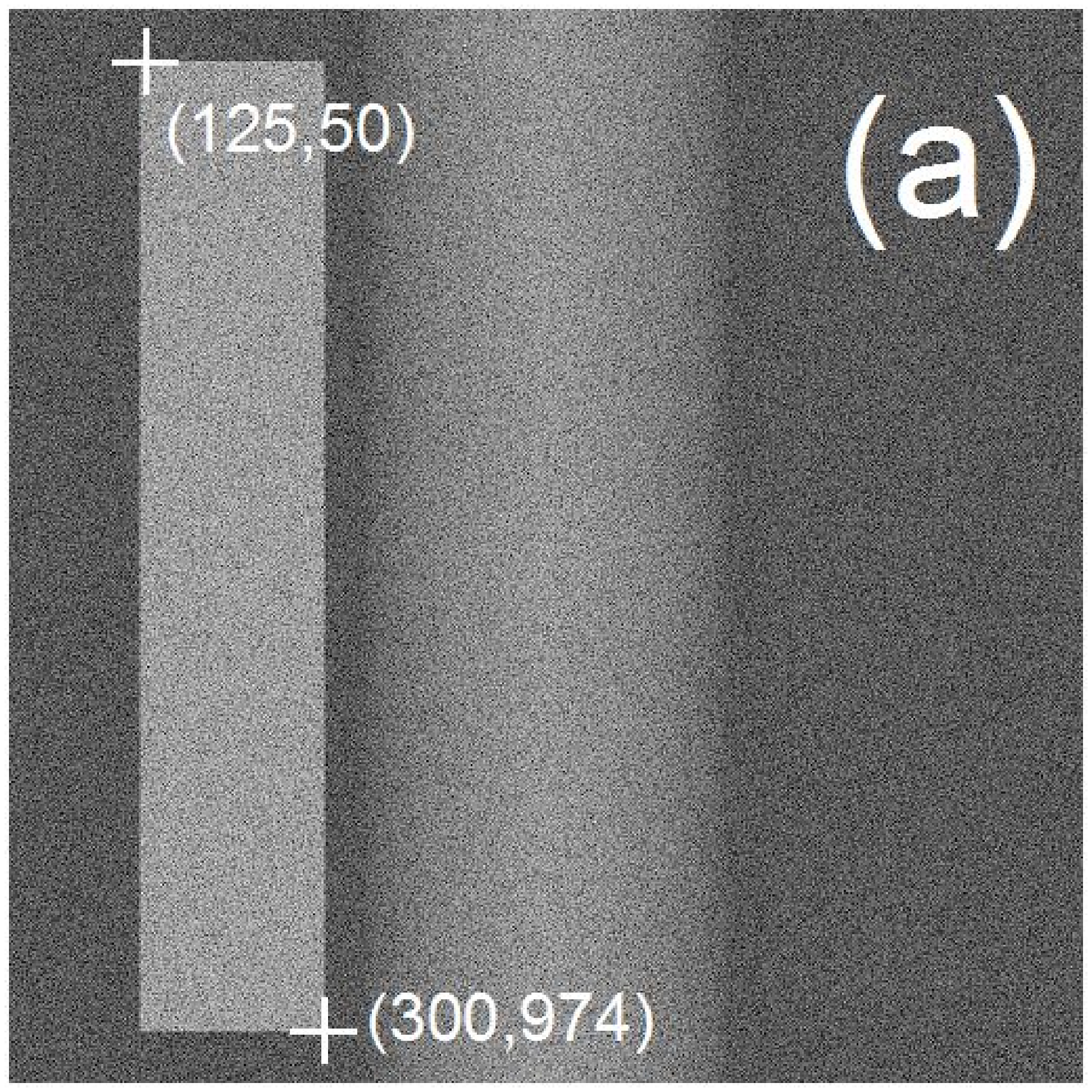}
 \includegraphics[width=4.2cm]{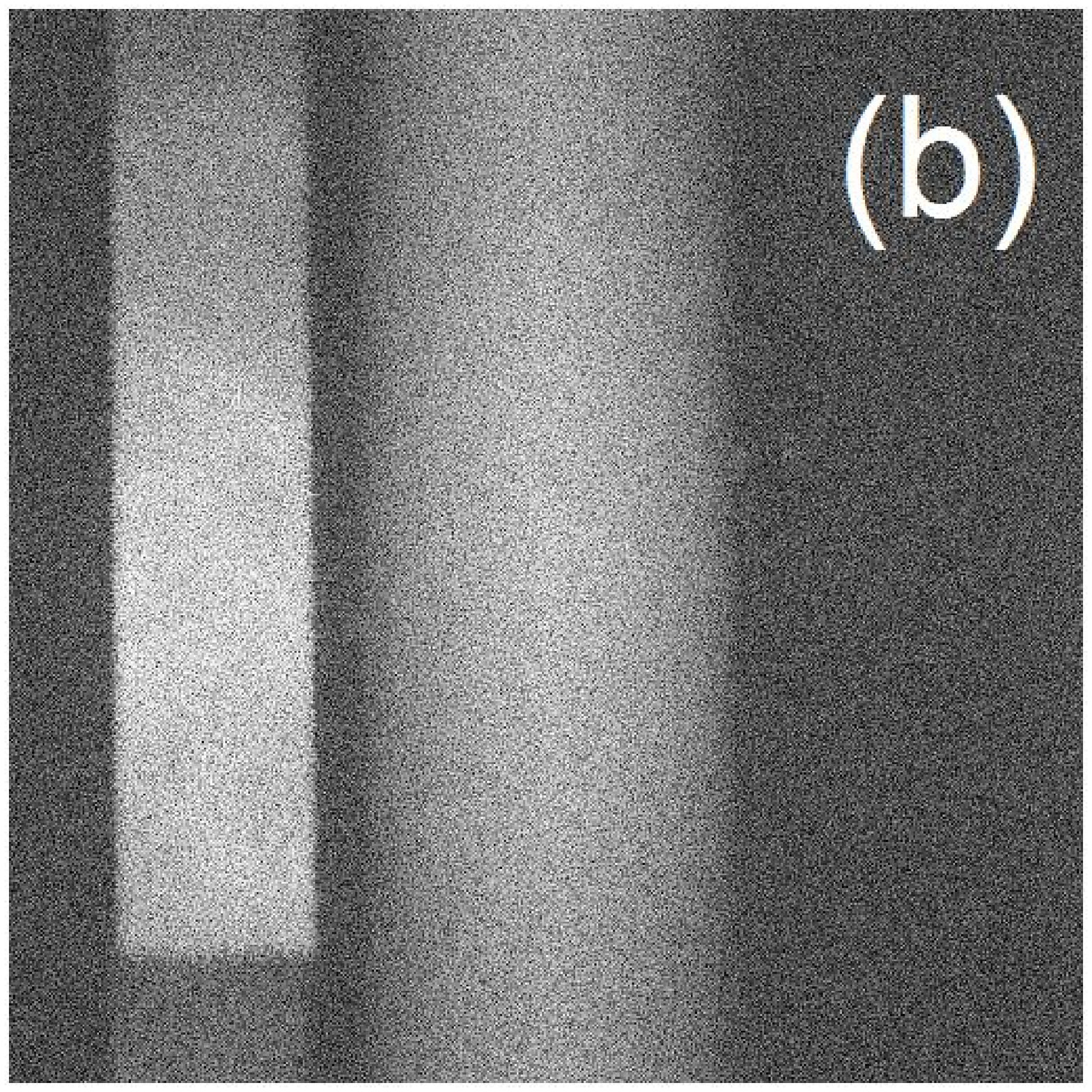}\\
   \includegraphics[width=4.2cm]{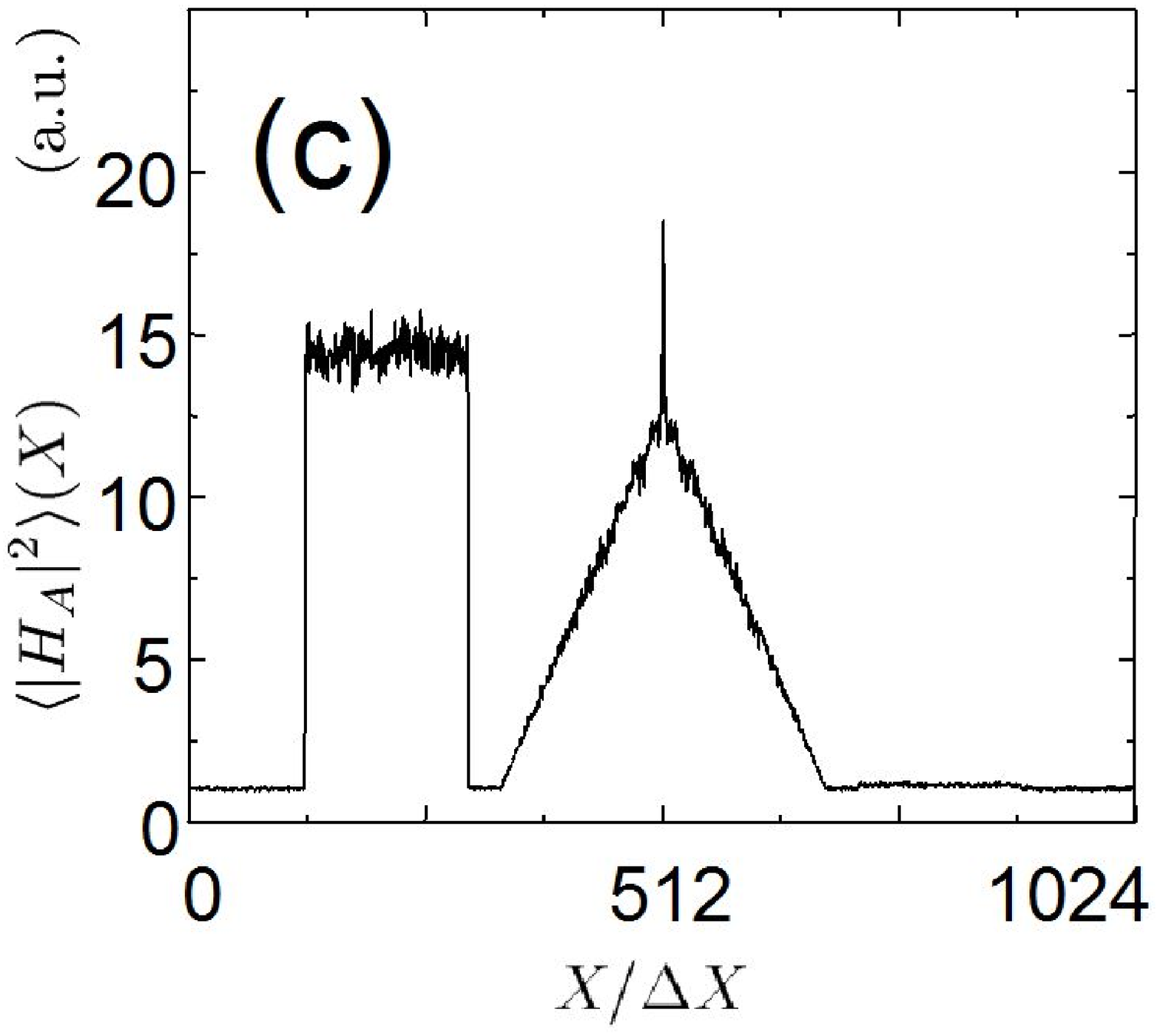}
  \includegraphics[width=4.2cm]{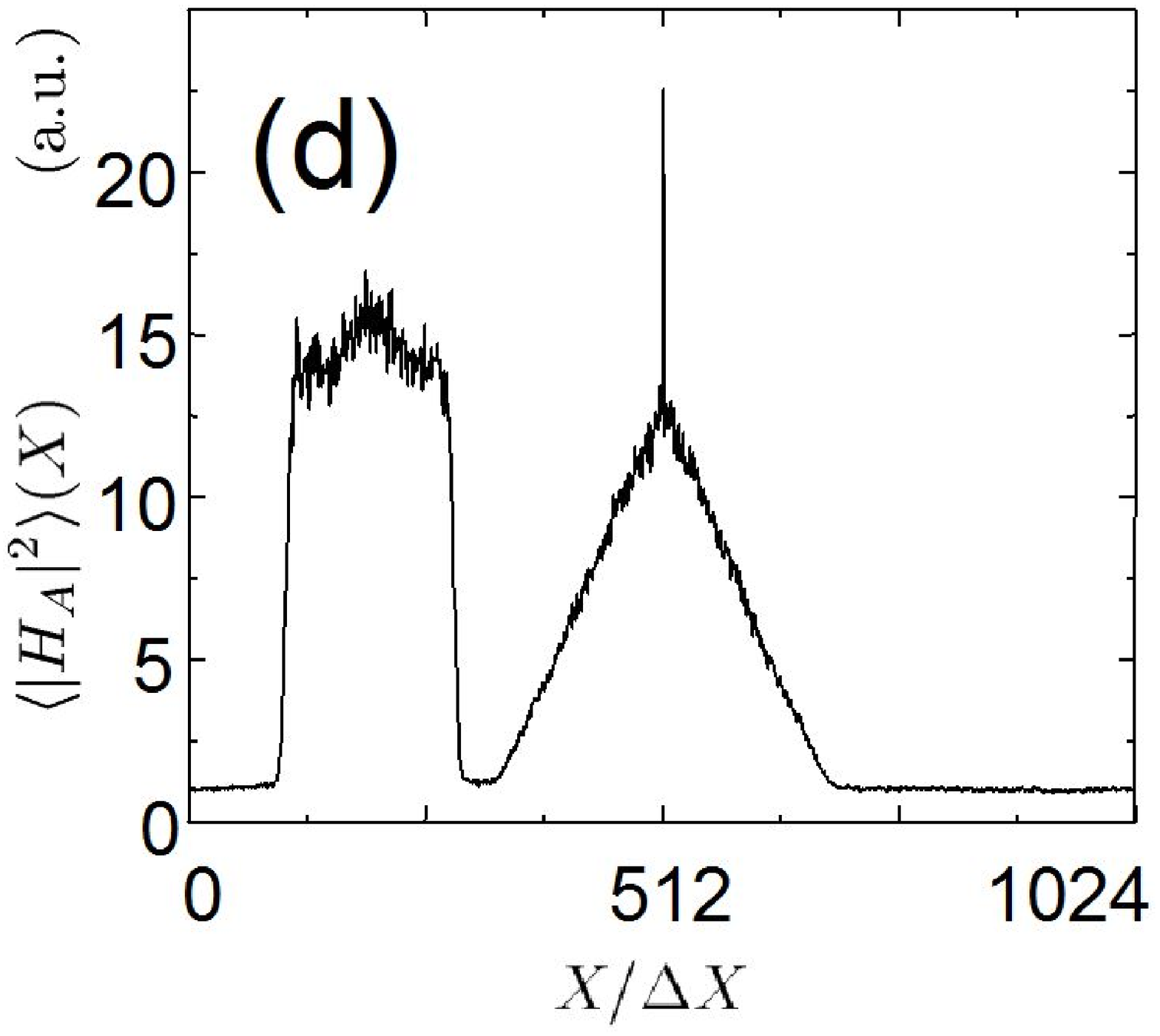}
  \caption{
Tagged intensity images $|H_{A}(X,Y)|^2$ (a,b) and curves $\langle
|H_{A}(X)|^2 \rangle$  (c,d)  obtained by calculation (a,c) and from
ref.\cite{gross2003shot} experiment (b,d). The images
$|H_{A}(X,Y)|^2$ (a,b) are displayed in an arbitrary logarithmic
scale. The curves $\langle  |H_{A}(X)|^2 \rangle$  (c,d) are
normalized  with respect to the ground floor. }\label{Fig_fig4}
\end{center}
\end{figure}

We have validated our theory by comparing the theoretical simulation
with the experiment of ref. \cite{gross2003shot}. Since the
experiment is made with low decorrelation, we have calculated
$H_{A}=H_{A,\textrm{corr}}$.
The calculation is made with $M=12$ frames, $|E_{LO}|^2= 10^4$ photo
electron per frames,  and   $\langle| E_{A,U}|^2 \rangle = 3.6\times
10^4$ and $\langle|E_{A,T}|^2 \rangle= 16$ photo electron per $T_C$
(12 frames). The size of the calculation grid is $1024\times 1024$.
The coordinates of the  upper left and bottom right aperture corners
are $(125, 50)$ and $(300, 974)$.

Figure \ref{Fig_fig4} shows   the arbitrary logarithmic scale
intensity image $|H_{A}(X,Y)|^2$ obtained by   calculation~(a), and
in experiment~(b)~\cite{gross2003shot}, and the corresponding
$\langle  |H_A|^2 \rangle(X)$ curves (c) and (d).
Like in ref.\cite{gross2003shot},   the curves were   normalized
with respect to  the noise floor  that corresponds to shot noise.
The maximum of tagged photon  signal $\langle  |H_A|^2 \rangle(X)$
is about $ 15$. This figure corresponds to  the tagged photon energy
for the whose sequence of $M=12$ frames: $\langle| E_{A,T}|^2
\rangle =16$.

The good   agreement of our calculation  with the experiment
validates our theoretical model.

\section{Conclusion}

In this paper, we have   proposed a theoretical model to describe
the detection of the tagged photons in heterodyne holography UOT.
This model, which  agrees with the results of \cite{gross2003shot},
has been used to calculate how   untagged photons, speckle noise,
shot noise, decorrelation and  etendue,  affect the UOT signal.

By a proper choice  of the aperture size,  heterodyne holography UOT
is able  to filter off the unwanted untagged photons, and to reach a
sensitivity limited by shot noise. This sensitivity    corresponds
to a noise   equivalent signal  equal to one  tagged  photo electron
per pixel (i.e. per etendue $\lambda^2$)  and per $T_C$, where
$T_C$ is either the recording time of the sequence (without
decorrelation) , or the exposure time of one frame (with
decorrelation).

We hope  this work will stimulate further  UOT development.

This work has been carried out  thanks to  the support of the LabEx
NUMEV project (n° ANR-10-LABX-20) funded by the "Investissements
d'Avenir" French Government program, managed by the French National
Research Agency (ANR)

\bibliographystyle{osajnl}


\end{document}